\documentclass[iop,twocolappendix,apj]{emulateapj}
\usepackage{natbib}
\usepackage{graphicx}
\usepackage{hyperref}
\usepackage{breakurl}
\usepackage{apjfonts}
\usepackage{longtable}
\usepackage{gensymb}
\usepackage{float}
\newcommand{\msun} {$M_{\odot}$}

\newcommand{\rsun} {$R_{\odot}$}

\newcommand{\hbeta} {H$\beta$}
\newcommand{\hgamma} {H$\gamma$}

\newcommand{\hepsilon} {H$\epsilon$}
\newcommand{\kms} {km s$^{-1}$}
\newcommand{\Te} {$T_{\rm eff}$}
\newcommand{\logg} {$\log g$}
\newcommand{\loghe} {$\log$ (He/H)}

\newcommand{\heib} {He {\sc i} $\lambda$4026}
\newcommand{\heia} {He {\sc i} $\lambda$4471}

\newcommand{\mgii} {Mg {\sc ii} $\lambda$5173-5184}
 
\newcommand{\caii} {Ca {\sc ii} $\lambda$3934}
\newcommand{\psr} {PSR~J1738+0333}
\newcommand{\lp} {LP~400-22}
\newcommand{\nltt} {NLTT~11748}

\begin{document}

\slugcomment{Accepted for publication in the Astrophysical Journal}
\shortauthors{GIANNINAS ET AL}
\shorttitle{THE ELM SURVEY VI}

\title{THE ELM SURVEY. VI. ELEVEN NEW DOUBLE DEGENERATES*}

\author{A. Gianninas$^{1, \dagger}$, Mukremin Kilic$^{1}$, Warren R. Brown$^{2}$, 
	Paul Canton$^{1, \dagger}$, and Scott J. Kenyon$^{2}$}

\affil{$^{1}$Homer L. Dodge Department of Physics and Astronomy,
  University of Oklahoma, 440~W.~Brooks~St., Norman, OK 73019, USA;
  alexg@nhn.ou.edu}
\affil{$^{2}$Smithsonian Astrophysical Observatory, 60~Garden~St.,
  Cambridge, MA 02138, USA}

\begin{abstract}
We present the discovery of 11 new double degenerate systems
containing extremely low-mass white dwarfs (ELM WDs). Our radial
velocity observations confirm that all of the targets have orbital
periods $\leq$~1 day. We perform spectroscopic fits and provide a
complete set of physical and binary parameters. We review and compare
recent evolutionary calculations and estimate that the systematic
uncertainty in our mass determinations due to differences in the
evolutionary models is small ($\approx$~0.01~\msun). Five of the new
systems will merge due to gravitational wave radiation within a Hubble
time, bringing the total number of merger systems found in the ELM
Survey to 38. We examine the ensemble properties of the current sample
of ELM WD binaries, including the period distribution as a function of
effective temperature, and the implications for the future evolution
of these systems. We also revisit the empirical boundaries of
instability strip of ELM WDs and identify new pulsating ELM WD
candidates. Finally, we consider the kinematic properties of our
sample of ELM WDs and estimate that a significant fraction of the WDs
from the ELM Survey are members of the Galactic halo.
\end{abstract}

\keywords{binaries: close -- Galaxy: stellar content -- gravitational
  waves -- supernovae: general -- techniques: spectroscopic -- white dwarfs}

\footnotetext[*]{Based on observations obtained at the MMT
  Observatory, a joint facility of the Smithsonian Institution and the
  University of Arizona.}
\footnotetext[$\dagger$]{Visiting Astronomer, Kitt Peak National Observatory, National Optical Astronomy Observatory, which is
operated by the Association of Universities for Research in Astronomy (AURA) under cooperative
agreement with the National Science Foundation.}

\section{INTRODUCTION}

Short period binary white dwarfs (WDs) are the proposed progenitors of
transient events such as supernovae Ia, underluminous .Ia, and Ca-rich
supernovae, and other exotic systems like AM CVn, R Coronae Borealis
(R CrB), and single subdwarf B/O stars
\citep{webbink84,iben84,bildsten07,perets10,solheim10,clayton13,foley15}. These
binary WDs, including the interacting AM CVn systems, are also
expected to be excellent gravitational wave sources and the only known
verification binaries for the evolved Laser Interferometer Space
Antenna \citep[$eLISA$,][]{amaro12}. The initial searches for short
period double WDs have found systems with periods as short as 1.4~h
\citep{moran97}, but they failed to find a large number of merging
binary systems \citep{napiwotzki07}.

Low-mass WDs with $M \sim 0.4$~\msun\ form $\approx10$\% of the WD
population in the solar neighborhood \citep{LBH05}, and the majority
of them are found in binary systems
\citep{marsh95,napiwotzki07,brownj11,debes15}. This is expected, as
the Galaxy is not old enough to produce $M\leq0.5$~\msun\ WDs through
single star evolution. The youngest WDs in Milky Way's globular
clusters have $M>0.5$~\msun\ \citep{hansen07}, consistent with this
explanation.

Extremely low-mass (ELM) WDs with $M<0.3$~\msun\ provide a unique
opportunity to significantly enlarge the known population of merging
WDs in the Galaxy. The ELM Survey is a targeted search for these WDs
in a well defined color and temperature range
\citep{brown_ELM1,brown_ELM3,brown_ELM5,kilic_ELM0,kilic_ELM2,kilic_ELM4}.
This survey has so far identified 56 binaries, all with $P\leq1$ d,
including 33 merger systems and four binaries with $P<1$ h. The two
shortest period systems, J0651 and J0935 (WD 0931+444), will merge in
$<$~1~Myr and $<$~10~Myr, respectively \citep{brown11,kilic14a}.

ELM WDs display a variety of photometric effects, including doppler
beaming \citep{shporer10}, tidal distortions, pulsations, and
eclipses. There are currently six eclipsing
\citep{steinfadt10,brown11,vennes11,parsons11,kilic14b,hallakoun15},
six pulsating \citep{hermes13c,kilic15}, and eight tidally distorted
\citep{hermes14} ELM WDs known. The discovery of new ELM WDs, hence,
provides new opportunities to improve the mass-radius relation and our
understanding of the interior structure of low-mass WDs.

\begin{figure*}[!ht]
\centering
\includegraphics[angle=-90,scale=0.515,bb=22 391 596 559]{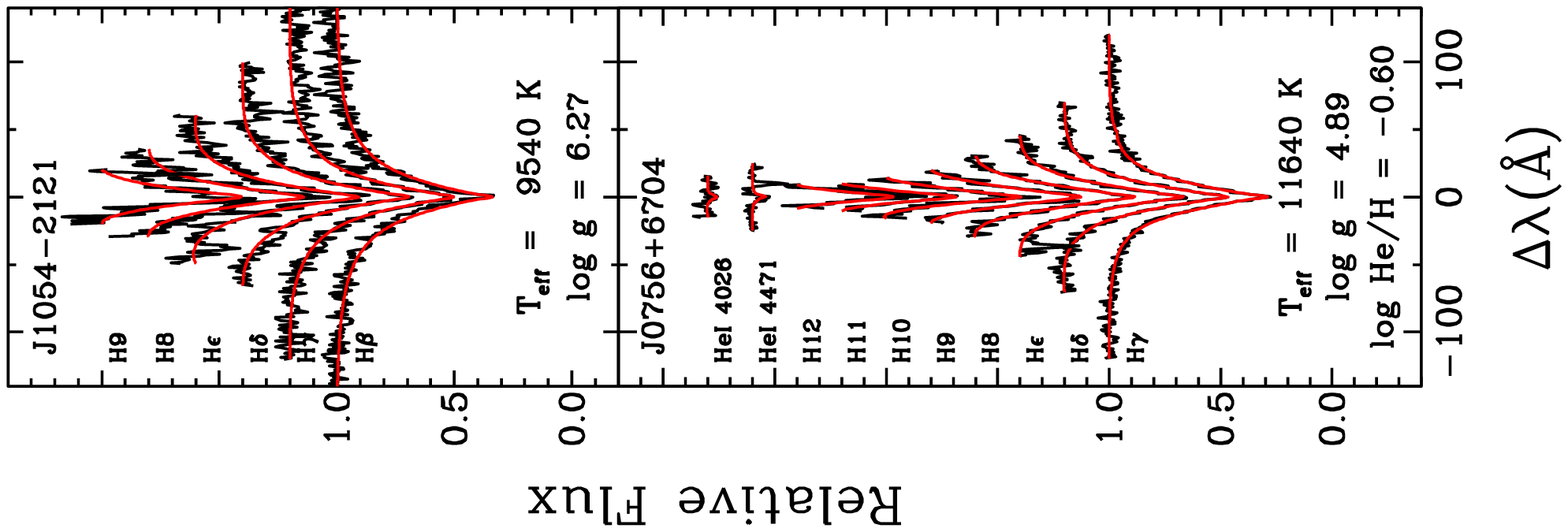}
\includegraphics[angle=-90,scale=0.515,bb=22 66 596 684]{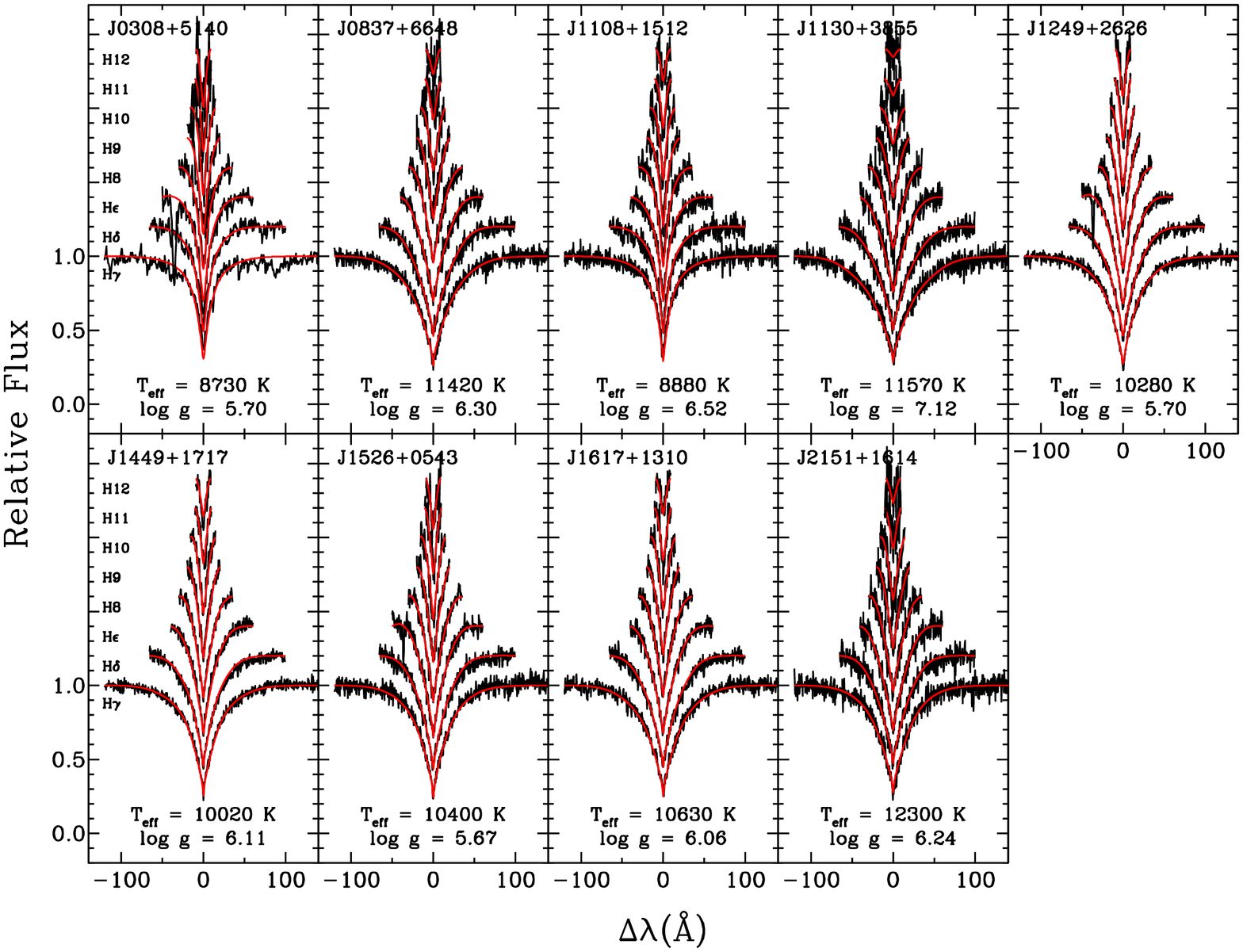}
\figcaption[f01a.eps]{1D model fits (red) to the observed Balmer line
  profiles (black) for the 11 new ELM WDs binaries. In the 
  right panel, we show fits to 9 ELM WDs, with optical
  spectra obtained at the MMT and FLWO, using pure H model atmospheres
  and include the lines from \hgamma\ (bottom) to H12 (top). We also
  show model fits for J1054$-$2121 (left panel, top) whose
  spectrum was obtained using the Kitt Peak 4~m and includes the lines
  from \hbeta\ to H9. Finally, fits using a mixed H-He model grid are
  shown for J0756+6704 (left panel, bottom) where we fit the
  \heib\ and \heia\ lines in addition to the Balmer lines. Individual
  spectral lines are offset by a factor of 0.2 for clarity. The
  best-fit atmospheric parameters are indicated at the bottom of each
  panel.
  \label{fg:fits}}
\end{figure*}

\citet{brown_ELM3} describe an efficient way to delineate ELM WDs from
normal WDs, A stars, and quasars, by targeting a color-selected sample
of $g=15-20$~mag low-mass WD candidates \citep[see Fig. 1
  of][]{brown_ELM3}.  Here we extend the target selection to the SDSS
Data Release 9 area while continuing to apply the same color selection
criteria and provide the first discoveries from this extended
survey. Three of our targets, J1130+3855, J1526+0543, and J1617+1310,
were previously identified as ELM WDs by \citet{brown_ELM3} based on
single epoch observations. Here we present follow-up radial velocity
observations of these three targets, as well as nine newly identified
ELM WDs.

We have also been obtaining radial-velocity measurements for
candidates from other sources including the Large Sky Area
Multi-Object Spectroscopy Telescope
\citep[LAMOST,][]{wang96,cui12}. \citet{zhao13} describe how a number
of WDs were identified from the LAMOST data but objects with
\logg\ $<$~7.0 were rejected as WDs. The spectra of these discarded
objects were kindly provided to us by J.~K.~Zhao (2013, private
communication). Based on our fits of these spectra, we identified a
number of possible ELM WD candidates including J0308+5140 and
J1249+2626. Coincidentally, J1249+2626 lies within the SDSS footprint,
however J0308+5140 does not.

Section~2 describes our spectroscopic observations. Section~3 presents
the physical and orbital parameters of the eleven new ELM WDs in
our survey. The sample characteristics of all 67 binaries in the
ELM Survey are discussed in Section~4 and we conclude in Section~5.

\section{OBSERVATIONS AND MODELLING}

\subsection{Optical Spectroscopy}

We used the 6.5m MMT telescope equipped with the Blue Channel
spectrograph \citep{schmidt89}, the 200-inch Hale telescope equipped
with the Double spectrograph \citep{oke82}, the Kitt Peak National
Observatory 4m telescope equipped with the R-C spectrograph, and more
recently with KOSMOS \citep{martini14}, to obtain spectroscopy of our
eleven targets in several observing runs. We operated the Blue
Channel, Double, R-C, and KOSMOS spectrographs with the
832~line~mm$^{-1}$, 1200~line~mm$^{-1}$, KPC22B, and b2k gratings,
providing wavelength coverages of 3650-4500~\AA, 3650-5180~\AA,
3700-5170~\AA\, and 3500-6200~\AA and spectral resolutions of 1.0~\AA,
1.7~\AA, 2.2~\AA, and 2.0~\AA, respectively. We also observed two of
the targets with $g<17$~mag in queue scheduled time at the 1.5~m FLWO
telescope using the FAST spectrograph \citep{fabricant98} with the
600~line~mm$^{-1}$ grating, providing 3500-5500~\AA\ wavelength
coverage with a spectral resolution of 2.3~\AA.  The Kitt Peak and
Palomar observations were obtained as part of the NOAO programs
2012A-0055, 2012B-0114, 2013A-0276, 2013B-0130, 2014A-0189, and
2015A-0082. The MMT and Hale observations were obtained at the
parallactic angle, but the Kitt Peak and FLWO observations were
obtained at a fixed slit angle. A comparison lamp exposure was
obtained with every observation. We flux-calibrate using blue
spectrophotometric standards \citep{massey88}.

We measure radial velocities using the cross-correlation package RVSAO
\citep{kurtz98}. Given the differences in wavelength coverage and
resolution of the five instruments used in our work, we
cross-correlate the observed spectra with the best-fit model WD
templates (see \S 2.2) at the appropriate resolution for each
spectrum. The average precision of our measurements ranges from about
10~\kms\ to 20~\kms. We compute best-fit orbital elements using the
code of \citet{kenyon86}, which weights each velocity measurement by
its associated error. We perform a Monte Carlo analysis to verify the
uncertainties in the orbital parameters \citep[see][]{brown_ELM3}.

\begin{figure}[!ht]
\includegraphics[scale=0.4,bb=12 124 592 708]{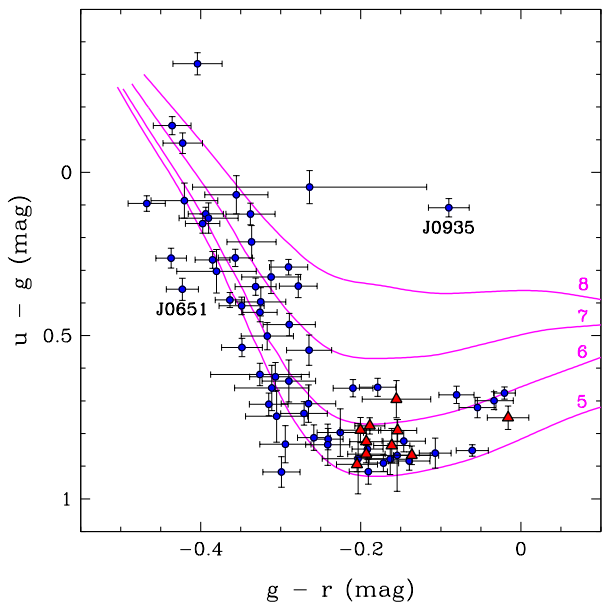}
\figcaption[f02.eps]{Color-color diagram of the WDs in the ELM Survey
  (blue circles), including 10 of the 11 systems
  identified in this paper (red triangles). Magenta lines show
  synthetic colors of WD model sequences with \Te\ = 30,000 -- 7500~K
  and \logg\ = 5, 6, 7, and 8. The two shortest period detached binary
  WD systems currently known, J0651 and J0935 (WD 0931+444), are
  labeled.
\label{fg:color}}
\end{figure}

\subsection{WD Model Atmosphere Analysis}

\begin{table*}\scriptsize
\caption{ELM WD Physical Parameters}
\begin{center}
\setlength{\tabcolsep}{6.5pt}
\begin{tabular*}{\hsize}{@{\extracolsep{\fill}}lccr@{ $\pm$ }@{\extracolsep{0pt}}lr@{ $\pm$ }@{\extracolsep{0pt}}lr@{ $\pm$ }@{\extracolsep{0pt}}lr@{ $\pm$ }@{\extracolsep{0pt}}lr@{ $\pm$ }@{\extracolsep{0pt}}lr@{ $\pm$ }@{\extracolsep{0pt}}lr@{ $\pm$ }@{\extracolsep{0pt}}l@{}}
\hline
\hline
\noalign{\smallskip}
SDSS & R.A.    & Decl.   & \multicolumn{2}{c}{\Te} & \multicolumn{2}{c}{\logg}         & \multicolumn{2}{c}{$M_{1}$} & \multicolumn{2}{c}{$g_{0}$} & \multicolumn{2}{c}{$M_{g}$} & \multicolumn{2}{c}{$d$} & \multicolumn{2}{c}{$\tau_{\rm cool}$} \\
     & (h:m:s) & (d:m:s) & \multicolumn{2}{c}{(K)} & \multicolumn{2}{c}{(cm s$^{-2}$)} & \multicolumn{2}{c}{(\msun)} & \multicolumn{2}{c}{(mag)} & \multicolumn{2}{c}{(mag)} & \multicolumn{2}{c}{(kpc)} & \multicolumn{2}{c}{(Gyr)}         \\
\noalign{\smallskip}
\hline
\noalign{\smallskip}
J0308+5140   & 03:08:18.19 &   +51:40:11.52 &  8380 & 140 & 5.51 & 0.10 & 0.151 & 0.024 & 13.049 & 0.010 &  8.04 & 0.53 & 0.100 & 0.025 & 1.128 & 0.358 \\
J0756+6704   & 07:56:10.71 &   +67:04:24.69 & 11640 & 250 & 4.90 & 0.14 & 0.181 & 0.011 & 16.233 & 0.020 &  5.22 & 0.46 & 1.597 & 0.343 & 0.189 & 0.078 \\
J0837+6648   & 08:37:08.51 &   +66:48:37.12 & 11400 & 240 & 6.31 & 0.05 & 0.181 & 0.011 & 17.844 & 0.019 &  8.83 & 0.25 & 0.634 & 0.072 & 1.157 & 0.123 \\
J1054$-$2121 & 10:54:35.78 & $-$21:21:55.94 &  9210 & 140 & 6.14 & 0.13 & 0.168 & 0.011 & 18.487 & 0.013 &  9.11 & 0.49 & 0.751 & 0.169 & 2.649 & 0.256 \\
J1108+1512   & 11:08:15.51 &   +15:12:46.74 &  8700 & 130 & 6.23 & 0.06 & 0.167 & 0.010 & 18.830 & 0.017 &  9.61 & 0.28 & 0.698 & 0.090 & 3.602 & 0.505 \\
J1130+3855   & 11:30:17.45 &   +38:55:50.11 & 11430 & 190 & 7.12 & 0.06 & 0.286 & 0.018 & 19.446 & 0.021 & 10.34 & 0.26 & 0.662 & 0.079 & 0.287 & 0.055 \\
J1249+2626   & 12:49:43.57 &   +26:26:04.22 & 10120 & 160 & 5.72 & 0.05 & 0.161 & 0.010 & 16.566 & 0.015 &  7.76 & 0.24 & 0.576 & 0.064 & 1.541 & 0.153 \\
J1449+1717   & 14:49:57.15 &   +17:17:29.33 &  9700 & 150 & 6.08 & 0.05 & 0.168 & 0.010 & 17.620 & 0.018 &  8.76 & 0.25 & 0.591 & 0.068 & 2.152 & 0.128 \\
J1526+0543   & 15:26:51.57 &   +05:43:35.31 & 10290 & 190 & 5.69 & 0.05 & 0.162 & 0.010 & 18.754 & 0.020 &  7.63 & 0.25 & 1.677 & 0.193 & 1.413 & 0.150 \\
J1617+1310   & 16:17:22.51 &   +13:10:18.87 & 10510 & 170 & 6.07 & 0.05 & 0.172 & 0.010 & 18.602 & 0.015 &  8.47 & 0.24 & 1.062 & 0.117 & 1.626 & 0.099 \\
J2151+1614   & 21:51:59.21 &   +16:14:48.72 & 12300 & 230 & 6.24 & 0.06 & 0.176 & 0.010 & 16.454 & 0.014 &  8.51 & 0.24 & 0.387 & 0.044 & 0.882 & 0.098 \\
\noalign{\smallskip}
\hline
\noalign{\smallskip}
\multicolumn{14}{@{}l}{{\bf Notes.} The listed \Te\ and \logg\ values are corrected for 3D effects according to \citet{tremblay15}.}\\
\end{tabular*}
\label{tab:par}
\end{center}
\end{table*}

Figure~\ref{fg:fits} presents the Balmer line profile fits for our
11 new ELM WDs. We use a pure-hydrogen model atmosphere grid
covering \Te\ = 4000 -- 35,000~K and \logg\ = 4.5 -- 9.5 and the
spectroscopic technique described in \citet[][and references
  therein]{gianninas11,gianninas14a} to fit the Balmer line profiles
of each target. The models include the Stark broadening profiles from
\citet{TB09}. We fit all of the visible Balmer lines up to H12 in the
MMT and Hale data, but we only fit \hbeta\ through H9 for the Kitt
Peak data due to the decrease in sensitivity below 3800~\AA. The
results from different telescopes agree within the errors, but we
adopt the values from the MMT data (when available) since those
spectra have the highest signal-to-noise ratio and resolution, and
include all of the higher order Balmer lines.

J0756+6704 also displays the \heib\ and \heia\ lines in its optical
spectrum. In this case,we use a grid of mixed H-He model atmospheres
to fit \Te, \logg, as well as the helium abundance. The measured
abundance of \loghe\ =~$-$0.60~$\pm$~0.38 coupled with the rather low
surface gravity of \logg =~4.89 makes it very similar to the mixed
H-He ELM WDs analyzed in \citet{gianninas14a} and is potentially a
sign of a recent shell flash.

We also note the continuing presence of the \caii\ line in the blue
wing of \hepsilon\ for all ELM WDs with \logg\ $<$~6.0
\citep{brown_ELM5,gianninas14a} including J0308+5140, J0756+6704,
J1249+2626, J1526+0543 as well as J1054$-$2121. Furthermore, in both
our spectrum and the SDSS spectrum of J0756+6704, the \mgii\ doublet
is clearly visible. The Na~D lines are also definitively detected in
the SDSS spectrum of J0756 and the LAMOST spectrum of
J0308+5140. These features mark J0756+6704 and J0308+5140 as most
intriguing systems worthy of a more detailed analysis
\citep[e.g.][]{gianninas14a,hermes14}. In all cases, the spectral
range where the Ca line is present is excluded from both the
normalization and fitting procedures.

Table~\ref{tab:par} summarizes the measured atmospheric parameters,
\Te\ and \logg. Note that the values listed in Table~\ref{tab:par}
have been corrected using the new 3D correction functions from
\citet{tremblay15} and thus differ from those displayed in
Figure~\ref{fg:fits}. The corrections have been applied to all targets
with \Te~$\leqslant$~12,000~K which includes all 11 new ELM WDs
save J2151+1614.

Based on the recent evolutionary sequences for ELM WDs by
\citet{althaus13}, we estimate masses ($M_{1}$), absolute magnitudes
($M_{g}$), and WD cooling ages ($\tau_{\rm cool}$) for each object.
Given $M_{g}$ and the extinction corrected SDSS $g$-band magnitude
($g_{0}$). However, since J0308+5140 is located outside the SDSS
footprint, for this object we estimate the $g$-band magnitude from The
Fourth US Naval Observatory CCD Astrograph Catalog \citep[UCAC4,
][]{zacharias13}. We also derive distances ($d$) to each object. These
are also included in Table \ref{tab:par}. All 11 targets have
\Te\ $\approx$~8000~--~12,000~K, \logg\ =~4.9~--~7.1,
$M$~$\approx$~0.15~--~0.29~\msun\ and they are located at
$d$~$\approx$~0.10~--~1.70~kpc from the Sun.

Contrary to previous analyses \citep{gianninas14a,gianninas14b}, we do
not adopt a systematic uncertainty of 0.02~\msun\ for our
determination of $M_{1}$. Instead, we propagate the uncertainties in
\Te\ and \logg\ through our interpolation of the \citet{althaus13}
model grid after which we add, in quadrature, the systematic
uncertainty of 0.01~\msun\ (see Section \ref{sec:evol}).

Figure~\ref{fg:color} shows a color-color diagram of all of the ELM
WDs identified in the ELM Survey, including 10 of the 11
systems described in this paper. The latter systems were selected to
have $g-r$~=~$-$0.2 to 0.0 mag based on the color-selection of
\citet{brown_ELM3}. Hence, it is not surprising that they all have
temperatures near 10,000 K. This figure demonstrates that ELM WDs have
colors consistent with the WD model sequences, except J0935 (WD
0931+444), whose photometry is contaminated by an M dwarf
\citep{kilic14a}.

\setcounter{footnote}{0}

\subsubsection{J0308+5140}

Fitting the optical spectrum of J0308+5140 required an additional
treatment. It is clear from the spectra that J0308+5140 suffers from a
considerable amount of reddening, consistent with its Galactic
latitude of $b$~=~$-$5.67$^{\circ}$. According to the NASA/IPAC
Infrared Science Archive (IRSA) dust
maps\footnote{http://irsa.ipac.caltech.edu/applications/DUST/}, the
reddening in the direction of J0308+5140 is $E(B-V)$~=~0.795.
Adopting this value for the reddening, we follow the prescription of
\citet{seaton79} to de-redden the observed spectrum of J0308+5140. The
slope of the corrected spectrum is in excellent agreement with the
slope of the spectroscopic solution. Consequently, we adopt an
extinction of $A_{g}$~=~3.014 from IRSA to correct the $g$-band
magnitude of J0308+5140 from UCAC4 (i.e. $g$~=~16.063) to get
$g_{0}$~=~13.049, as listed in Table~\ref{tab:par}.

\begin{figure}[!t]
\includegraphics[scale=0.435,bb=18 144 592 698]{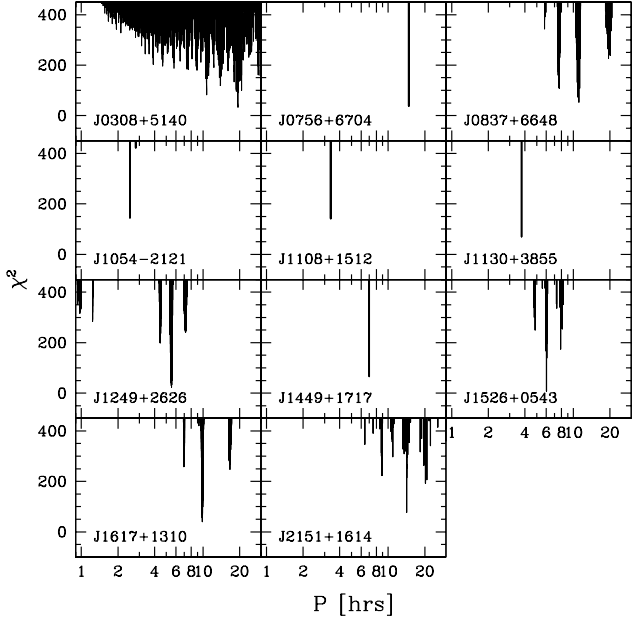}
\figcaption[f03.eps]{Periodograms for the 11 new ELM WD
  binaries. The best orbital periods have the smallest $\chi^{2}$
  values; some binaries are well constrained and some have period
  aliases.
\label{fg:chi2}}
\end{figure}

\begin{figure}[!ht]
\begin{minipage}[][][t]{0.475\textwidth}
\centering
\includegraphics[scale=0.415,bb=18 144 592 698]{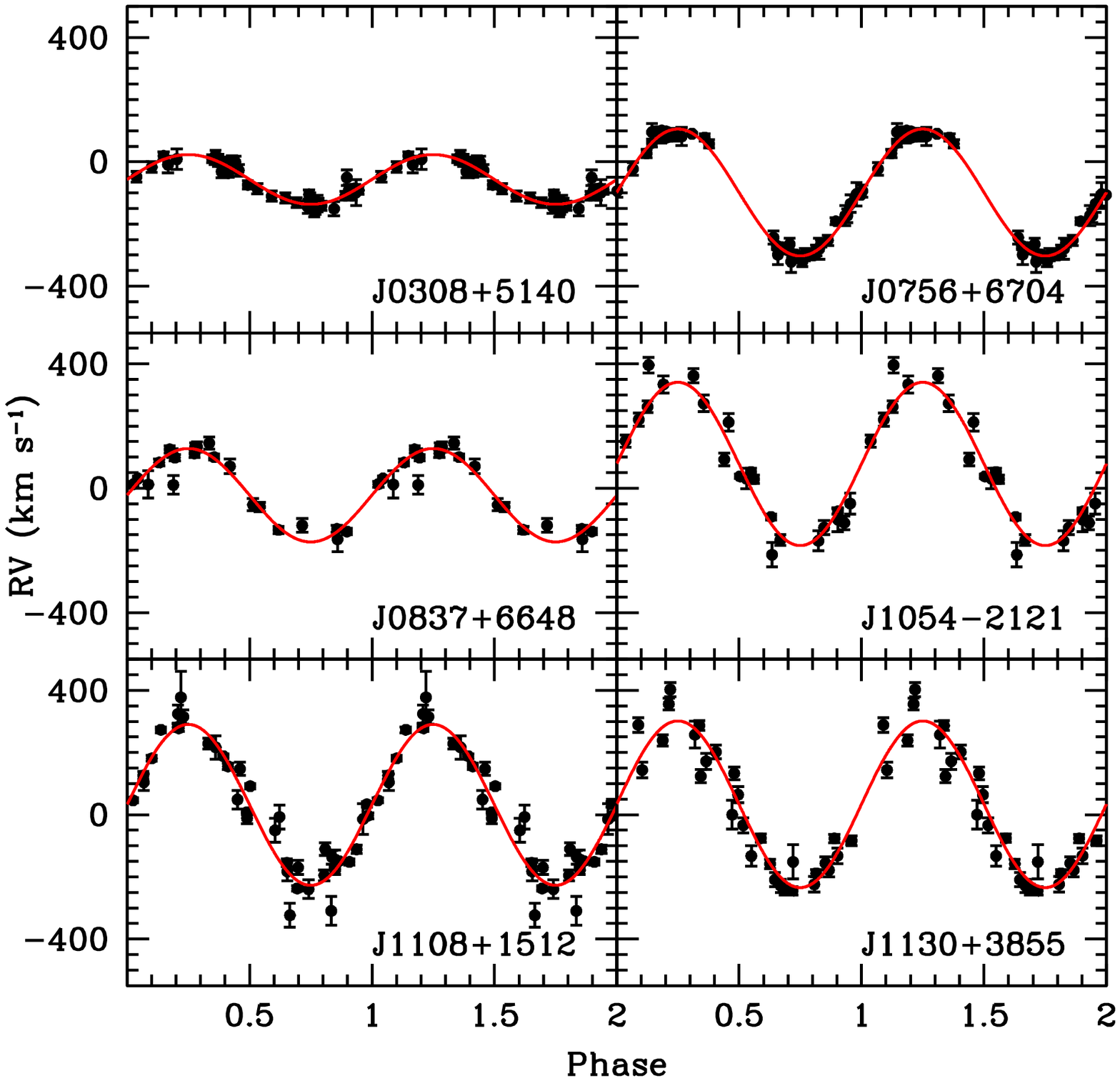}
\par\vfill
\includegraphics[scale=0.415,bb=18 144 592 698]{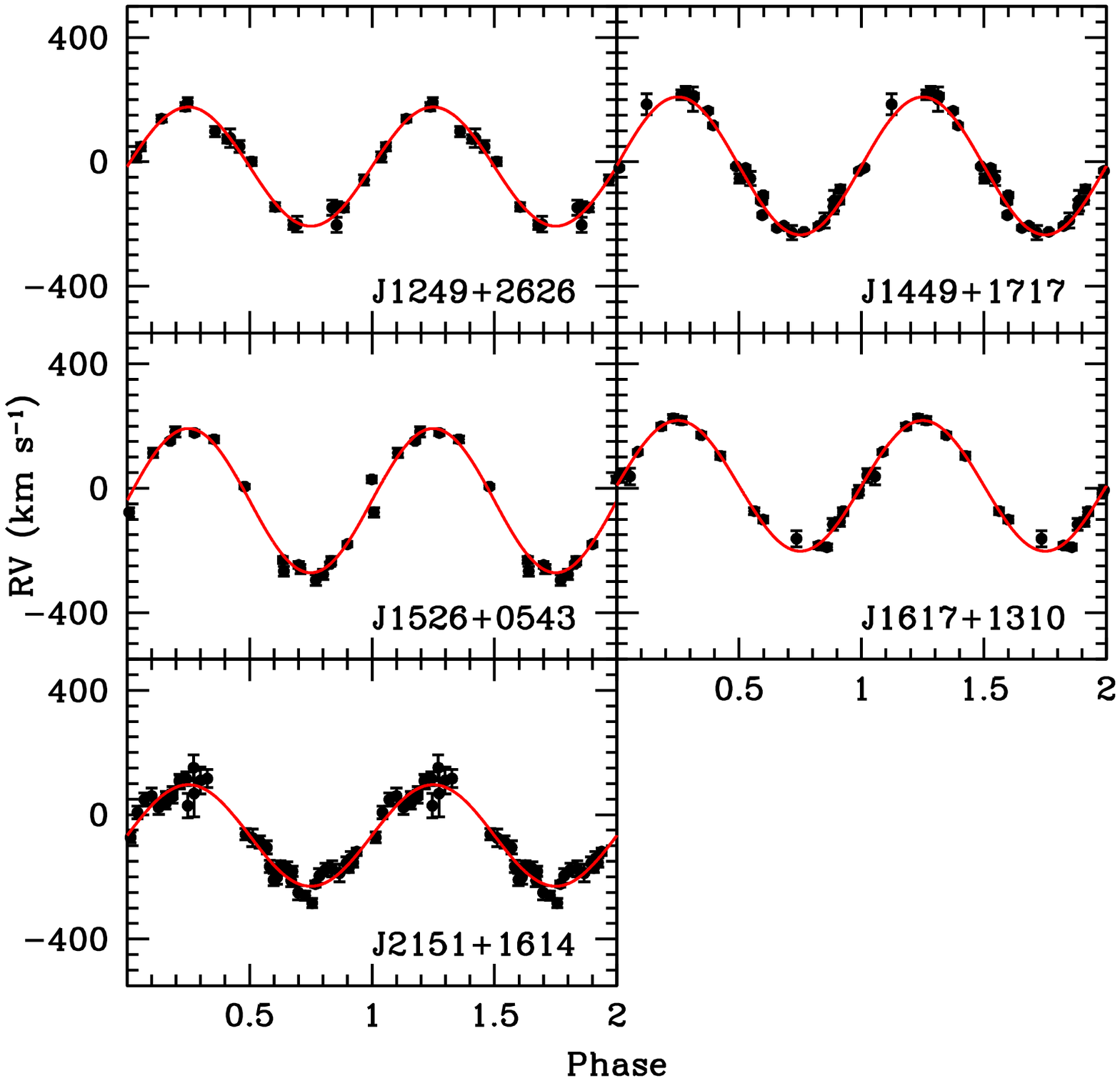}
\figcaption[fo4a.eps]{Observed velocities phased to the best-fit
  orbits (red) for the 11 new ELM WD binaries.
\label{fg:rv}}
\end{minipage}
\end{figure}

\begin{table*}\footnotesize
\caption{ELM WD Binary Parameters}
\begin{center}
\setlength{\tabcolsep}{5.5pt}
\begin{tabular*}{\hsize}{@{\extracolsep{\fill}}lr@{ $\pm$ }@{\extracolsep{0pt}}lr@{ $\pm$ }@{\extracolsep{0pt}}lr@{ $\pm$ }@{\extracolsep{0pt}}lr@{ $\pm$ }@{\extracolsep{0pt}}lr@{ $\pm$ }@{\extracolsep{0pt}}lr@{ $\pm$ }@{\extracolsep{0pt}}lcr@{ $\pm$ }@{\extracolsep{0pt}}lc@{}}
\hline
\hline
\noalign{\smallskip}
SDSS & \multicolumn{2}{c}{$P$} & \multicolumn{2}{c}{$K$} & \multicolumn{2}{c}{$\gamma$} & \multicolumn{2}{c}{Mass Function} & \multicolumn{2}{c}{$M_{2}$} & \multicolumn{2}{c}{$M_{2, i=60\degree}$} & $\tau_{\rm merge}$ & \multicolumn{2}{c}{$a$} & $\log h$ \\
     & \multicolumn{2}{c}{(days)} & \multicolumn{2}{c}{(\kms)} & \multicolumn{2}{c}{(\kms)} & \multicolumn{2}{c}{(\msun)} & \multicolumn{2}{c}{(\msun)} & \multicolumn{2}{c}{(\msun)} & (Gyr) & \multicolumn{2}{c}{(\rsun)} & \\
\noalign{\smallskip}
\hline
\noalign{\smallskip}
J0308+5140   & 0.80590 & 0.00038 &  78.9 & 2.7 &  $-$56.8 & 1.8 & 0.041 & 0.004 & $\geqslant$~0.16 & 0.02 & 0.20 & 0.03 & \ldots             & 2.46 & 0.12 & $-$22.57 \\
J0756+6704   & 0.61781 & 0.00002 & 204.2 & 1.6 &  $-$98.7 & 1.3 & 0.545 & 0.013 & $\geqslant$~0.82 & 0.03 & 1.13 & 0.04 & \ldots             & 3.05 & 0.04 & $-$23.05 \\
J0837+6648   & 0.46329 & 0.00005 & 150.3 & 3.0 &  $-$22.4 & 2.3 & 0.163 & 0.010 & $\geqslant$~0.37 & 0.02 & 0.48 & 0.03 & \ldots             & 2.06 & 0.04 & $-$22.85 \\
J1054$-$2121 & 0.10439 & 0.00655 & 261.1 & 7.1 &     94.3 & 4.5 & 0.193 & 0.028 & $\geqslant$~0.39 & 0.05 & 0.52 & 0.06 & $\leqslant$~1.452  & 0.77 & 0.06 & $-$22.49 \\
J1108+1512   & 0.12310 & 0.00867 & 256.2 & 3.7 &     44.5 & 2.4 & 0.214 & 0.024 & $\geqslant$~0.42 & 0.04 & 0.56 & 0.05 & $\leqslant$~2.143  & 0.87 & 0.07 & $-$22.48 \\
J1130+3855   & 0.15652 & 0.00001 & 284.0 & 4.9 &     24.5 & 3.6 & 0.371 & 0.019 & $\geqslant$~0.72 & 0.04 & 0.96 & 0.05 & $\leqslant$~1.652  & 1.23 & 0.02 & $-$22.14 \\
J1249+2626   & 0.22906 & 0.00112 & 191.6 & 3.9 &  $-$15.5 & 2.4 & 0.167 & 0.011 & $\geqslant$~0.35 & 0.02 & 0.47 & 0.03 & $\leqslant$~13.242 & 1.26 & 0.03 & $-$22.65 \\
J1449+1717   & 0.29075 & 0.00001 & 228.5 & 3.2 &      3.9 & 3.0 & 0.359 & 0.015 & $\geqslant$~0.59 & 0.03 & 0.81 & 0.04 & \ldots             & 1.69 & 0.03 & $-$22.54 \\
J1526+0543   & 0.25039 & 0.00001 & 231.9 & 2.3 &  $-$39.8 & 1.8 & 0.324 & 0.010 & $\geqslant$~0.54 & 0.02 & 0.74 & 0.02 & $\leqslant$~12.087 & 1.49 & 0.02 & $-$22.99 \\
J1617+1310   & 0.41124 & 0.00086 & 210.1 & 2.8 &      7.7 & 2.0 & 0.395 & 0.017 & $\geqslant$~0.64 & 0.03 & 0.87 & 0.04 & \ldots             & 2.17 & 0.04 & $-$22.86 \\
J2151+1614   & 0.59152 & 0.00008 & 163.3 & 3.1 &  $-$67.0 & 2.3 & 0.267 & 0.015 & $\geqslant$~0.49 & 0.03 & 0.66 & 0.04 & \ldots             & 2.59 & 0.05 & $-$22.61 \\
\noalign{\smallskip}
\hline
\noalign{\smallskip}
\end{tabular*}
\label{tab:bin}
\end{center}
\end{table*}

\section{ORBITAL PARAMETERS}

Figure~\ref{fg:chi2} shows the periodograms for our 11 targets.
All of our targets have a well defined minimum $\chi^{2}$ in the
periodograms and the period is constrained to $P\leq$~1 d in all
cases. Furthermore, to obtain a $p$-value of $p$~=~0.01 with four
degrees of freedom requires a $\Delta \chi^{2} \geq 13.3$, with
respect to the minimum $\chi^{2}$ \citep{press86}. This criterion is
met by all the new ELM WD binaries. Thus, the orbital periods we
determine are statistically significant at the 99\% confidence level.

Table~\ref{tab:bin} and Figure~\ref{fg:rv} present the best-fit
orbital parameters and the radial velocity curves,
respectively. Table~\ref{tab:bin} lists the orbital period ($P$),
radial velocity semi-amplitude ($K$), systemic velocity ($\gamma$),
the mass function, the minimum secondary mass ($M_{2}$) assuming
$i$~=~90$^{\circ}$, the secondary mass assuming $i$~=~60$^{\circ}$,
the maximum gravitational wave merger time ($\tau_{\rm merge}$), the
orbital separation ($a$) and finally, the gravitational wave strain
($\log h$).

Our targets show radial velocity variations with orbital periods of $P
\approx$~3--19~h and semi-amplitudes of $K \approx 80$~to $\approx
285$~\kms. The systemic velocities range from $\gamma \approx -$100 to
$\approx$~100~\kms. We note that the systemic velocities do not
include the corrections for the WDs' gravitational redshift, which is
a few \kms.

\begin{figure}[!t]
\centering
\includegraphics[scale=0.425,bb=20 117 592 684]{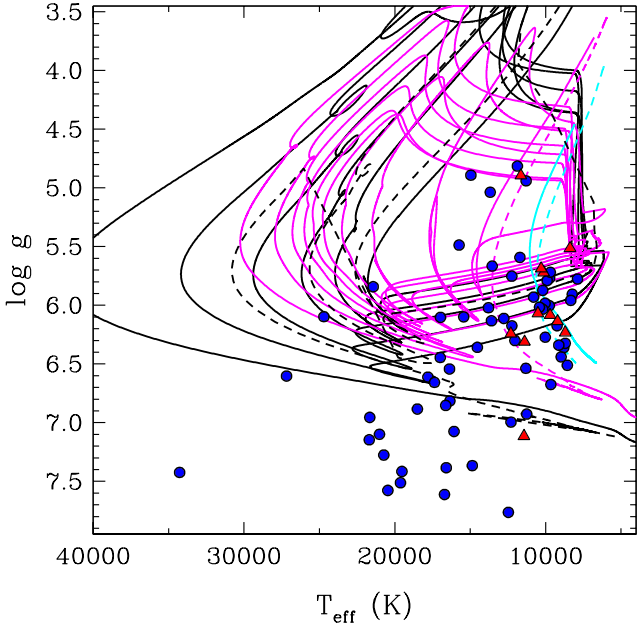}
\figcaption[f05.eps]{Location in the \Te\ -- \logg\ plane of the
  11 new ELM WDs presented in this paper (red triangles) along with
  the complete ELM Survey sample (blue circles). Also plotted are
  evolutionary tracks from \citet{althaus13} (solid lines) and
  \citet{istrate14} (dashed lines) for 0.165~\msun\ (cyan),
  0.187~\msun\ (magenta), and 0.239~\msun\ (black).
\label{fg:tg}}
\end{figure}

\begin{figure}[!t]
\centering
\includegraphics[scale=0.425,bb=25 67 592 684]{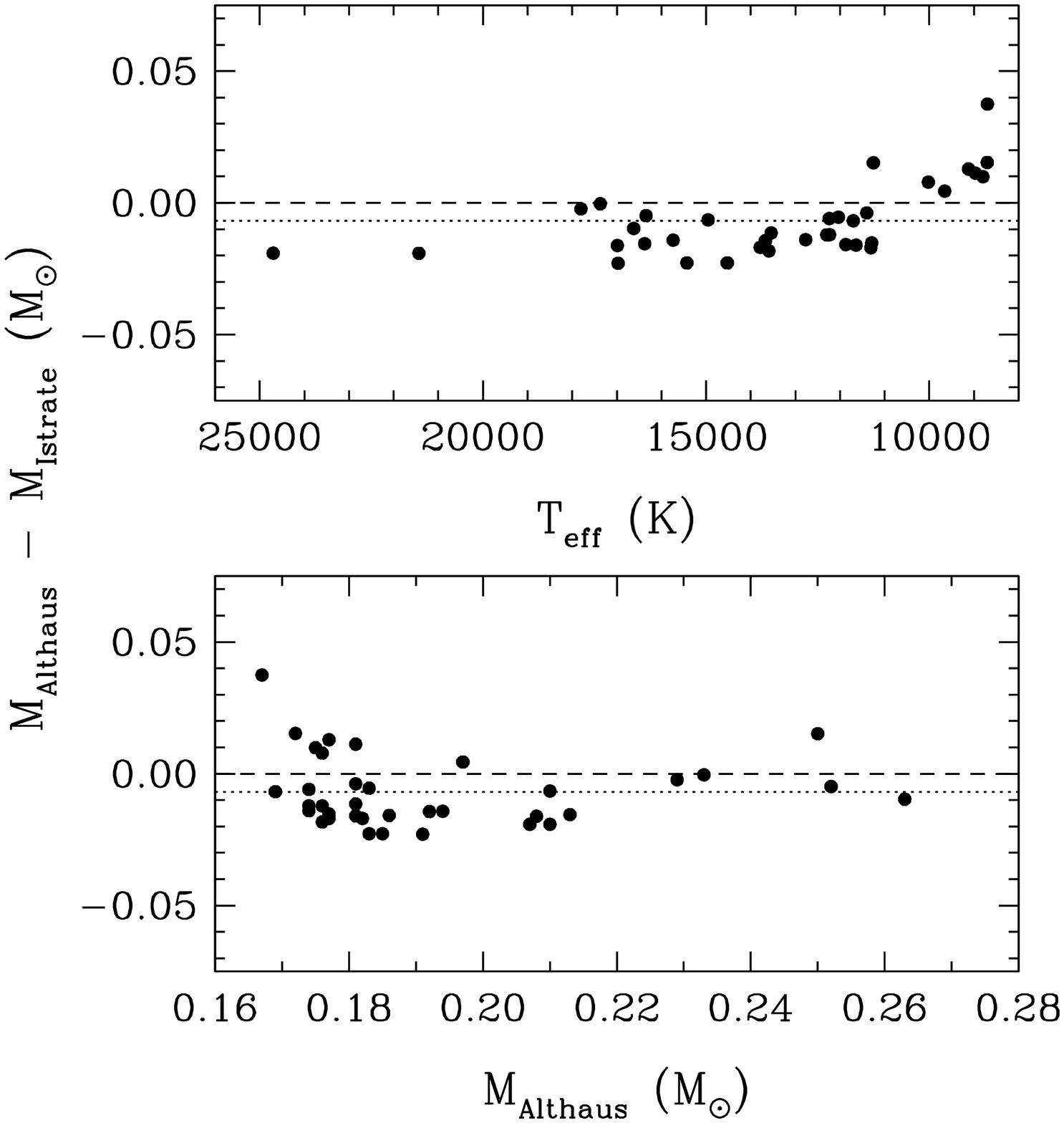}
\figcaption[f06.eps]{Difference in the primary mass as determined
  using the models from \citet[][$M_{\rm Althaus}$]{althaus13} and
  \citet[][$M_{\rm Istrate}$]{istrate14} plotted as a function of
  \Te\ (top) and $M_{\rm Althaus}$ (bottom). The dotted line denotes
  the average difference.
\label{fg:delM}}
\end{figure}

The minimum mass companions to these 11 ELM WDs range from
0.16~\msun\ to 0.82~\msun. Based on the mass function, the probability
of a 1.4--3.0~\msun\ neutron star companion ranges from 2\% to a
maximum of 20\% for the companion of J0756+6704. These percentages are
quite low and it is unlikely that any of these 11 new ELM WDs
have neutron star companions. These probabilities agree with the
recent studies of \citet{andrews14} and \citet{boffin15} whose results
imply that neutron star companions to ELM WDs should be rare.

It is also unlikely that the total masses for any of these new systems
is above the Chandrasekhar mass limit. Assuming an average inclination
of $i$~=~60$^{\circ}$, the binaries with the highest total system mass
are J0756+6704 and J1130+3855 with $M_{\rm tot}$~=~1.31~\msun\ and
1.25~\msun, respectively.

Our sample also includes five new merger systems (J1054$-$2121,
J1108+1512, J1130+3855, J1249+2626, J1526+0543). The quickest to merge
will be J1054$-$2121 with $\tau_{\rm merge}$~=~1.452~Gyr. However, the
gravitational wave strains of these systems are not strong enough to
be detected by $eLISA$ \citep[see Figure 12 of][]{gianninas14a}.

\section{DISCUSSION}

\subsection{Eleven New Binary WD Systems}

With these 11 new discoveries, and including the pulsating ELM
WDs published by \citet{hermes13b,hermes13c}, and the discoveries
presented in \citet{kilic14a,kilic14b} and \citet{gianninas14b}, the
ELM Survey has found a total of 73 ELM WDs of which 67 are
in detached, double-degenerate binaries; 38 of the binaries will merge
within a Hubble time. Counting only six systems where no significant
radial velocity variability has been detected, 92\% of the ELM Survey
targets are formally members of compact ($P < 1$~d) binary systems.
In addition to the results listed in Tables~\ref{tab:par} and
\ref{tab:bin}, we provide, for completeness, the physical and binary
parameters for the entire ELM Survey sample in Tables~\ref{tab:par2}
and \ref{tab:bin2} of the Appendix. Note that for ELM WDs with
\Te~$<$~12,000~K, the atmospheric parameters have been adjusted using
the 3D corrections from \citet{tremblay15}.
\\
\subsection{The ELM WD Sample}

\subsubsection{Comparison of Evolutionary Models}\label{sec:evol}

In Table~\ref{tab:par} we adopt masses derived from the evolutionary
sequences for He-core WDs of \citet{althaus13}. However, new
evolutionary sequences appropriate for He-core ELM WDs have recently
been published by \citet{istrate14}. Due to the importance of
obtaining accurate masses in order to correctly predict the final
merger products of ELM WD binaries, it is important to explore these
new evolutionary calculations.

One of the major differences between the \citet{althaus13} and
\citet{istrate14} models are the initial assumptions with regards to
the progenitors and companions of the eventual ELM WDs. All of the
\citet{althaus13} models assume a 1.0~\msun\ progenitor for the ELM WD
and are evolved in a binary with a 1.4~\msun\ neutron star
companion. In contrast, the \citet{istrate14} models assume a range of
progenitor masses (from 1.2 to 1.6~\msun) as well as a range of
neutron star masses (from 1.3 to 1.75 \msun) to produce the full range
of ELM WD masses in their model grid. As we will demonstrate, both
approaches yield similar ELM WD masses. However, as we mentioned in
Section 3, neutron star companions are unlikely in most of these
systems. Consequently, it would be useful to have evolutionary tracks
computed for ELM WD binaries with massive WD companions instead to
provide additional insight into the final ELM WD masses.

\begin{figure}[!t]
\centering
\includegraphics[scale=0.4,bb=45 117 592 679]{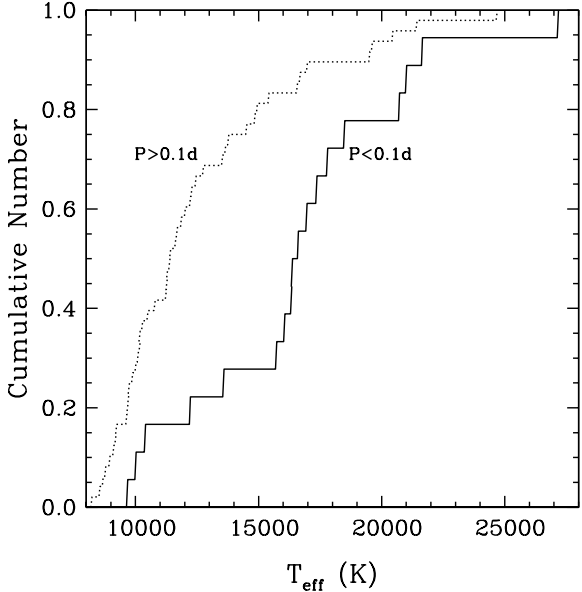}
\figcaption[f07.eps]{Cumulative distribution as a function of \Te\ for
  ELM WDs with $P < 0.1$~d (solid line) and $P > 0.1$~d (dotted line).
\label{fg:histoTP}}
\end{figure}

\begin{figure}[!t]
\centering
\includegraphics[scale=0.4,bb=45 117 592 679]{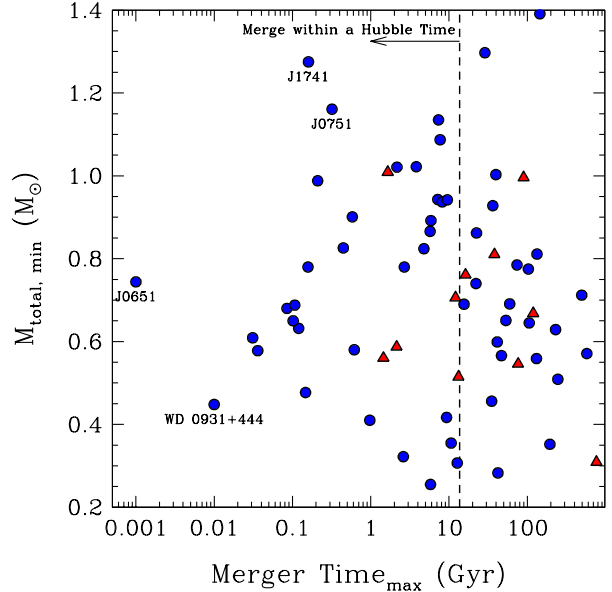}
\figcaption[f08.eps]{Plot of the minimum total mass as a function of
  the maximum merger time for the entire ELM Survey sample (blue
  circles) with the 11 new binaries from this paper (red
  triangles).
\label{fg:merge}}
\end{figure}

In Figure~\ref{fg:tg}, we plot the location of all ELM WDs from the
ELM Survey sample in the \Te\ -- \logg\ plane. In addition, we plot a
selection of evolutionary tracks from both \citet{althaus13} and
\citet{istrate14}. In order to make the most direct comparison
possible, we have chosen a set of three representative evolutionary
tracks corresponding to final ELM WD masses of 0.165~\msun,
0.187~\msun, and 0.239~\msun.

The results in Figure~\ref{fg:tg} demonstrate that there are
significant differences between the two sets of evolutionary
tracks. Most notably, the \citet{istrate14} models predict
considerably fewer shell flashes than the \citet{althaus13}
tracks. This is most clearly seen in the 0.187~\msun\ tracks (magenta)
where the Istrate model predicts no shell flashes while the equivalent
Althaus model undergoes a series of shell flashes. Indeed, the Althaus
models assume $Z$~=0.01 while the Istrate models assume
$Z$~=0.02. Since the maximum mass for which flashes occur increases
with lower metallicity \citep{nelson04}, it is not surprising that the
Althaus models predict a greater number of shell flashes and for a
broader range of masses.

Despite these significant differences, the most important question is
whether or not the models yield the same masses for a given \Te\ and
\logg. We have chosen a set of nine models from the grid of
\citet{istrate14} to compute masses. To greatly simplify the
calculation, we consider only the final cooling portion of the
evolutionary tracks.  In Figure~\ref{fg:delM} we plot the difference
in masses derived from the two model grids as function of \Te\ and
$M_{\rm Althaus}$.  Figure~\ref{fg:delM} demonstrates that the Althaus
\& Istrate models yield very similar masses. The average difference is
$\approx$~0.01~\msun\ with an overall scatter of roughly the same
magnitude. Consequently, we adopt an uncertainty of 0.01~\msun\ as the
systematic error due to uncertainties in the evolutionary models.
This is equivalent to an error of $\approx$~5\% in mass. In contrast,
the mean error in mass due to uncertainties in the atmospheric
parameters, \Te\ and \logg, is $\approx$~0.006~\msun\ which is
$\approx$~3\%.  Therefore, the error in our mass estimates is
dominated by the systematic uncertainties from the evolutionary
models.

\subsubsection{The Period Distribution of Binary WDs}

In Figure~\ref{fg:histoTP} we plot the cumulative distributions, as a
function of \Te, for the 18 systems with P~$<$~0.1~d and the 49
systems with P~$>$~0.1~d. We see that the majority of systems with
P~$>$~0.1~d are found with \Te\ $\lesssim$~15,000~K.

To determine if the two distributions are statistically independent,
we perform the two-sample Kolmogorov-Smirnov (K-S) test. The K-S
statistic, $D_{n, n'}$, is defined as the largest difference between
the cumulative distributions functions and can be expressed as

\begin{displaymath}
D_{n, n'} = \sup_{x} |F_{1,n}(x) - F_{2,n'}(x)|
\end{displaymath}

\noindent where $F_{1,n}(x)$ and $F_{2,n'}(x)$ represent the $P
>$~0.1~d and $P <$~0.1~d distribtuions, respectively. For the given
distributions in Figure~\ref{fg:histoTP}, we determine a K-S statistic
of $D_{n, n'}$~=~0.5590 and a $p$-value of $p$~=~0.0002. Based on this 
result, we can reject the null hypothesis at
the $\alpha$~=~0.01 significance level. However, the K-S test is not
always sensitive enough to determine if two distributions are indeed
independent. This is because, by definition, the distributions
converge to 0 and 1 at the ends. The Anderson-Darling (A-D) test is
more useful in such situations. We perform the A-D test and obtain a
standardized test statistic of $T$~=~6.9807 and $p$~=~0.0006. Based on
these values, we once again reject the null hypothesis at the
$\alpha$~=~0.01 significance level. The results of these statistical
tests strongly suggests that the two samples are \emph{not} drawn from
the same parent distribution.

The fact that the two samples appear to originate from separate parent
distributions can be understood intuitively. The shorter period
systems ($P < 0.1$~d) will merge before they have a chance to cool to
\Te~$\lesssim$~15,000~K. On the other hand, the longer period binaries
($P > 0.1$~d) have maximum merger times which are considerably longer
and therefore they can survive long enough to cool below
\Te~$\lesssim$~15,000~K.

\begin{figure}[!t]
\centering
\includegraphics[scale=0.325,angle=-90,bb=37 16 596 784]{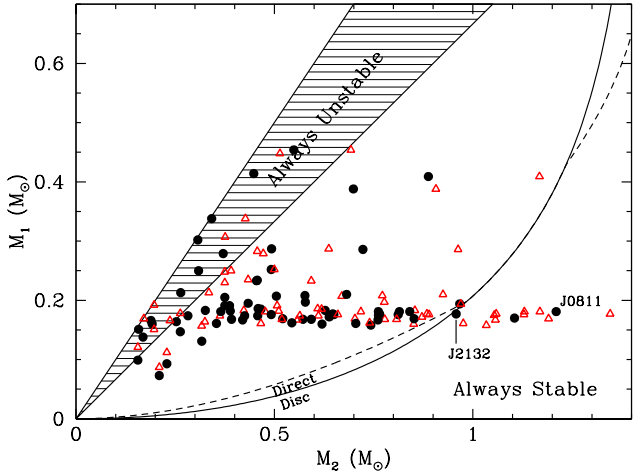}
\figcaption[f09.eps]{Plot of $M_{1}$ vs. $M_{2}$ for the entire ELM
  sample. Black points represent $M_{2,{\rm min}}$ while the red
  triangles assume $M_{2,i=60\degree}$. The majority of the ELM WD
  binaries lie in the region between the regions of stable and
  unstable mass transfer. \label{fg:stable}}
\end{figure}

\begin{figure}[!t]
\centering
\includegraphics[scale=0.425,bb=20 67 592 679]{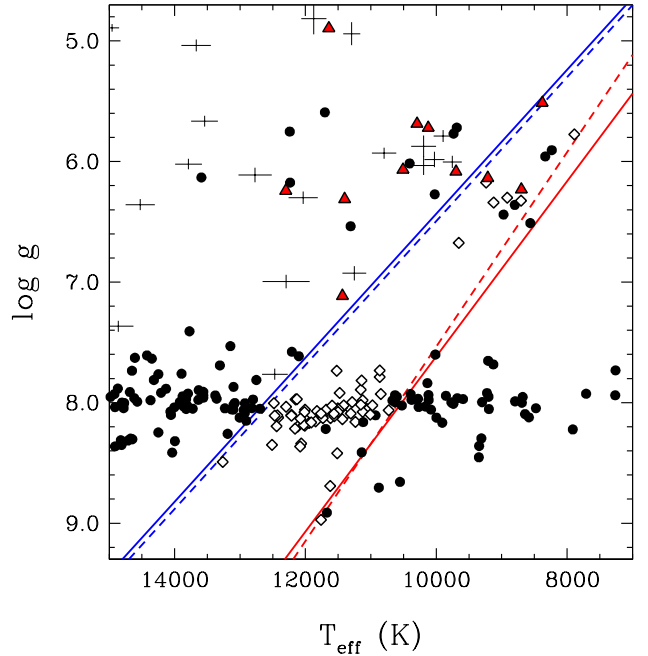}
\figcaption[f10.eps]{Region of the \Te\ -- \logg\ plane containing the
  canonical ZZ Ceti instability strip (lower left) and the six
  currently known ELM pulsators (upper right). All pulsators are
  identified as open diamonds, whereas WDs that have been confirmed as
  photometrically constant are represented as filled dots. The solid
  lines denote the blue and red boundaries from
  \citep{gianninas14a}. The blue and red dashed lines indicate the
  tentative new boundaries which take into account the 3D corrected
  values of \Te\ and \logg. Error bars denote WDs from the ELM Survey
  sample which have not yet been investigated for photometric
  variability. Red triangles indicate the 11 new ELM WDs
  binaries.
\label{fg:ZZ}}
\end{figure}

\subsubsection{The Future: Merger Products}

In Figure~\ref{fg:merge} we plot the total minimum system mass
(assuming $i$~=~90$^{\circ}$) as a function of the maximum merger time
of the system. The maximum gravitational wave merger time is given by

\begin{equation}
\tau_{\rm merge,max} = \frac{(M_{1}+M_{2})^{1/3}}{M_{1}M_{2}}P^{8/3} \times 10^{-2} {\rm Gyr}
\end{equation}

\noindent where $M_{1}$ is the mass of the ELM WD in \msun, $M_{2}$ is
the minimum companion mass in \msun, and the period $P$ is in hours
\citep{landau58}.

Our new sample of 11 ELM WD binaries has yielded five new
systems (J1054$-$2121, J1108+1512, J1130+3855, J1249+2626, J1526+0543)
which are expected to merge within a Hubble time. Thus, the ELM Survey
has now identified a total of 38 merger systems.

J0651 and WD~0931+444 remain the two shortest period systems, expected
to merge in $\approx$~1~Myr and $\approx$~10~Myr, respectively. The
most massive merger systems J1741+6526, J0751$-$0141, near the top of
Figure~\ref{fg:merge}, represent potential Type Ia
progenitors. However, the total mass of the system is not the only
parameter that determines the ultimate fate of the
binary. \citet{marsh04} showed that systems with high mass ratios will
in fact lead to stable mass transfer binaries.  In
Figure~\ref{fg:stable} we plot the stability diagram from
\citet{marsh04} and also plot the values of $M_{1}$ and $M_{2}$ for
the entire ELM Survey sample for both the minimum and most likely
(i.e. for $i$~=~60$^{\circ}$) companion mass. Note that in all cases,
we choose $M_{2} > M_{1}$.

As \citet{kilic14b} have shown, J1741+6526 and J0751$-$0141 are, given
the uncertainties in $M_{2}$, located in the region of the stability
diagram which predicts that they will eventually become stable mass
transfer systems. Thus, they represent the progenitors of future AM
CVn systems. However, Figure~\ref{fg:stable} also shows that two more
ELM WD binaries lie this same region, namely J0811+0225 and
J2132+0754. This brings the total to four unambiguous identifications
of AM CVn progenitor systems. In addition, a number of systems do lie
in the the unstable regime, which predicts an eventual
merger. However, the total mass for all of these merger systems is
significantly below the Chandrasekhar mass.

Figure~\ref{fg:stable} also clearly shows that most ELM WD binaries
are found in the regime between the stable and unstable
regions. Indeed, there is still much uncertainty and debate over the
eventual fate of compact double WD binaries. The recent calculations
of \citet{shen15} predict that most double WD binaries will eventually
merge. On the other hand, \citet{kremer15} predict that a majority of
systems will undergo stable mass transfer. Once the ELM Survey is
completed, a more rigorous examination of both the \citet{shen15} and
\citet{kremer15} models will be possible by comparing these merging
ELM WD binaries with the known populations of various merger products
including AM~CVn systems, R CrB stars and Type Ia supernovae.

\subsubsection{Instability Strip}

The discovery of the first five pulsating ELM WDs were reported in
\citet{hermes13c} and references therein. Recently, \citet{kilic15}
discovered the sixth pulsating ELM WD which also happens to be the
companion to \psr\ \citep{antoniadis12}. These pulsating ELM WDs all
have \Te\ = 8000 -- 10,000~K and \logg\ = 6 -- 7 \citep[see the
  updated instability strip in Figure~13 of][]{gianninas14a}.
Furthermore, the location of this newly defined instability strip is
in good agreement with the predictions of pulsation models
\citep{vg13}. In Figure~\ref{fg:ZZ} we plot an updated version of the
figure from Gianninas et al. including the 11 new ELM WDs from
this paper and the canonical ZZ Ceti instability strip
\citep{gianninas11} including several new discoveries presented by
\citet{green15}. One important difference here is that the adopted
\Te\ and \logg\ values for all the WDs in the figure have been
corrected for 3D effects using the functions defined in
\citet{tremblay13b} for WDs with \logg~$>$~7.0 and \citet{tremblay15}
for the ELM WD regime. In particular, the corrected atmospheric
parameters for \psr\ are \Te~=~8910~$\pm$~150~K and
\logg~=~6.30~$\pm$~0.10.

Our best-fit temperatures and surface gravities place five of
our new ELM WDs within (J1054$-$2121, J1108+1512) or near the blue
edge (J0308+5140, J1130+3855, J1449+1717) of the instability strip as
delineated by our new boundaries defined by Equations~\ref{eq:blue}
and \ref{eq:red}, respectively. Follow-up high-speed photometric
observations of these five targets will be useful to search for
pulsations in these ELM WDs and are under way with the McDonald 2.1m
telescope (K. Bell 2015, private communication).

\begin{equation}
(\log g)_{\rm blue} = 5.96923 \times 10^{-4} (T_{\rm eff})_{\rm blue} + 0.52431
\label{eq:blue}
\end{equation}

\begin{equation}
(\log g)_{\rm red} = 8.06630 \times 10^{-4} (T_{\rm eff})_{\rm red} - 0.53039
\label{eq:red}
\end{equation}

\subsubsection{Disk vs. Halo Membership}

\begin{table}[!t]
\caption{Space Velocities of ELM WDs}
\begin{center}
\begin{tabular*}{\hsize}{@{\extracolsep{\fill}}lr@{ $\pm$ }@{\extracolsep{0pt}}rr@{ $\pm$ }@{\extracolsep{0pt}}rr@{ $\pm$ }@{\extracolsep{0pt}}rcc@{}}
\hline
\hline
\noalign{\smallskip}
SDSS & \multicolumn{2}{c}{$U$}    & \multicolumn{2}{c}{$V$}    & \multicolumn{2}{c}{$W$}    & $D_{\rm m,disk}$ & $D_{\rm m,halo}$ \\
     & \multicolumn{2}{c}{(\kms)} & \multicolumn{2}{c}{(\kms)} & \multicolumn{2}{c}{(\kms)} &                  &                  \\
\noalign{\smallskip}
\hline
\noalign{\smallskip}
J0022$-$1014 &     82 & 23 &  $-$40 &  21 &     27 &  8 & ~~1.944 & 1.505 \\
J0056$-$0611 &   $-$4 &  7 &  $-$42 &   9 &  $-$11 &  3 & ~~0.520 & 1.376 \\
J0106$-$1000 & $-$126 & 51 & $-$236 &  55 &  $-$16 & 16 & ~~5.173 & 1.122 \\
J0112+1835   &     67 &  8 & $-$104 &   9 &     52 &  8 & ~~2.677 & 1.055 \\
J0152+0749   &     82 & 12 & $-$117 &  19 &  $-$15 & 13 & ~~2.597 & 0.812 \\
J0345+1748   & $-$147 &  6 & $-$197 &  25 &  $-$28 &  4 & ~~4.889 & 1.194 \\
J0651+2844   &   $-$8 &  4 &      2 &  15 &   $-$9 & 15 & ~~0.545 & 1.795 \\
J0745+1949   &  $-$67 &  4 &  $-$10 &   4 &     23 &  4 & ~~1.729 & 1.797 \\
J0755+4800   &  $-$23 &  3 &     25 &   2 &     31 &  3 & ~~1.440 & 2.053 \\
J0815+2309   &    153 & 19 & $-$275 &  38 &   $-$6 & 25 & ~~6.037 & 1.270 \\
J0822+2753   &     64 &  6 &  $-$20 &   9 &  $-$24 &  7 & ~~1.434 & 1.622 \\
J0825+1152   &  $-$71 & 18 &     54 &  24 &  $-$13 & 24 & ~~2.217 & 2.359 \\
J0849+0445   &  $-$20 & 15 &  $-$20 &  16 &     25 & 17 & ~~0.957 & 1.630 \\
J0900+0234   &   $-$8 & 14 &  $-$52 &  13 &     38 & 14 & ~~1.363 & 1.368 \\
J0917+4638   &  $-$25 & 22 &     19 &  32 &      9 & 23 & ~~1.067 & 1.973 \\
J0923+3028   &     11 &  3 &  $-$16 &   4 &      2 &  3 & ~~0.231 & 1.618 \\
J1046$-$0153 &     10 &  6 &   $-$4 &   7 &  $-$51 &  6 & ~~1.427 & 1.797 \\
J1053+5200   & $-$112 & 22 & $-$201 &  29 &     15 & 13 & ~~4.440 & 0.947 \\
J1056+6536   &     20 & 22 &  $-$27 &  24 &     20 & 18 & ~~0.769 & 1.532 \\
J1104+0918   &  $-$23 &  3 &  $-$84 &   6 &     31 &  5 & ~~1.716 & 1.080 \\
J1141+3850   &    280 & 91 & $-$345 & 102 &    106 & 32 & ~~9.378 & 2.664 \\
J1234$-$0228 &   $-$2 & 12 &  $-$70 &  12 &     25 &  6 & ~~1.292 & 1.157 \\
J1238+1946   &     64 & 47 & $-$652 &  65 &  $-$92 &  9 & 12.952 & 4.492 \\
J1422+4352   &  $-$48 & 39 &  $-$72 &  35 & $-$165 & 18 & ~~4.879 & 2.071 \\
J1436+5010   &     43 & 13 &      9 &  12 &  $-$26 &  8 & ~~1.228 & 1.871 \\
J1443+1509   &  $-$33 &  9 & $-$167 &  22 & $-$141 &  5 & ~~4.994 & 1.505 \\
J1448+1342   &      6 & 10 &      2 &  11 &  $-$21 &  7 & ~~0.687 & 1.795 \\
J1512+2615   &   $-$5 & 11 &  $-$60 &  13 &   $-$5 &  7 & ~~0.824 & 1.209 \\
J1518+0658   &  $-$38 &  4 &  $-$21 &   5 &  $-$35 &  3 & ~~1.288 & 1.645 \\
J1538+0252   & $-$179 & 17 & $-$126 &  24 &   $-$1 & 19 & ~~4.496 & 1.503 \\
J1557+2823   &     26 &  3 &      5 &   3 &     17 &  3 & ~~0.889 & 1.824 \\
J1625+3632   &     56 & 57 & $-$136 &  49 &  $-$22 & 44 & ~~2.639 & 0.581 \\
J1630+2712   &     36 & 36 &  $-$91 &  34 & $-$106 & 30 & ~~3.332 & 1.411 \\
J1630+4233   &     34 & 13 &   $-$8 &  10 &  $-$10 & 10 & ~~0.727 & 1.691 \\
J1741+6526   &     39 & 14 &  $-$55 &   8 &  $-$14 & 12 & ~~1.072 & 1.261 \\
J1840+6423   &    118 & 18 &  $-$61 &   6 &      2 & 12 & ~~2.617 & 1.389 \\
J2132+0754   &     48 & 14 &  $-$32 &  10 &   $-$6 & 13 & ~~0.993 & 1.474 \\
J2228+3623   &     20 &  5 &  $-$37 &   5 &     26 &  5 & ~~0.947 & 1.457 \\
J2236+2232   & $-$328 & 33 & $-$181 &  11 &   $-$9 & 13 & ~~7.903 & 2.447 \\
J2338$-$2052 &     18 & 22 &     11 &  22 &      3 &  8 & ~~0.702 & 1.868 \\
J2345$-$0102 &    343 & 35 & $-$206 &  26 &     69 & 15 & ~~8.507 & 2.450 \\
\noalign{\smallskip}
\hline
\noalign{\smallskip}
\multicolumn{9}{c}{This paper}\\
\noalign{\smallskip}
\hline
\noalign{\smallskip}
J0308+5140   &     55 &  1 &  $-$24 &   2 &     12 &  2 & ~~1.189 & 1.574 \\
J0756+6704   &     17 & 21 & $-$227 &  48 &     29 & 24 & ~~4.244 & 0.523 \\
J0837+6648   &      4 &  7 &  $-$53 &  10 &      8 &  7 & ~~0.727 & 1.276 \\
J1054$-$2121 &      8 & 11 &  $-$64 &   7 &     63 &  9 & ~~2.080 & 1.369 \\
J1108+1512   &  $-$29 & 11 &  $-$40 &  11 &     24 &  6 & ~~1.121 & 1.454 \\
J1130+3855   &  $-$85 & 15 &  $-$29 &  12 &   $-$1 &  6 & ~~1.951 & 1.658 \\
J1526+0543   &     21 & 21 &      2 &  27 &  $-$54 & 19 & ~~1.575 & 1.855 \\
J2151+1614   &  $-$53 &  7 &  $-$44 &   3 &     14 &  5 & ~~1.405 & 1.452 \\
\noalign{\smallskip}
\hline
\end{tabular*}
\label{tab:UVW}
\end{center}
\end{table}

\begin{figure}[!t]
\centering
\includegraphics[scale=0.425,bb=20 167 592 679]{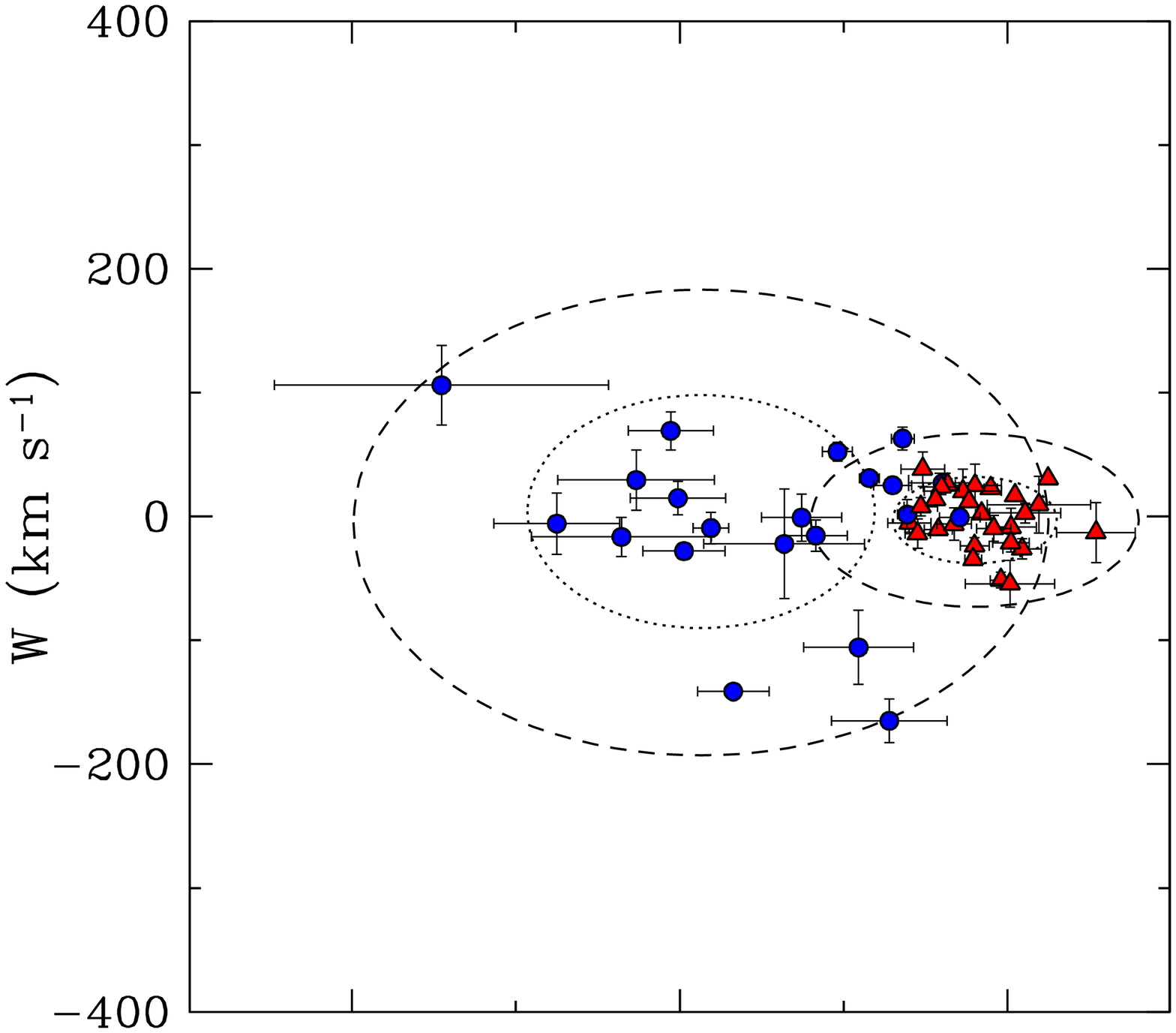}
\includegraphics[scale=0.425,bb=20 117 592 679]{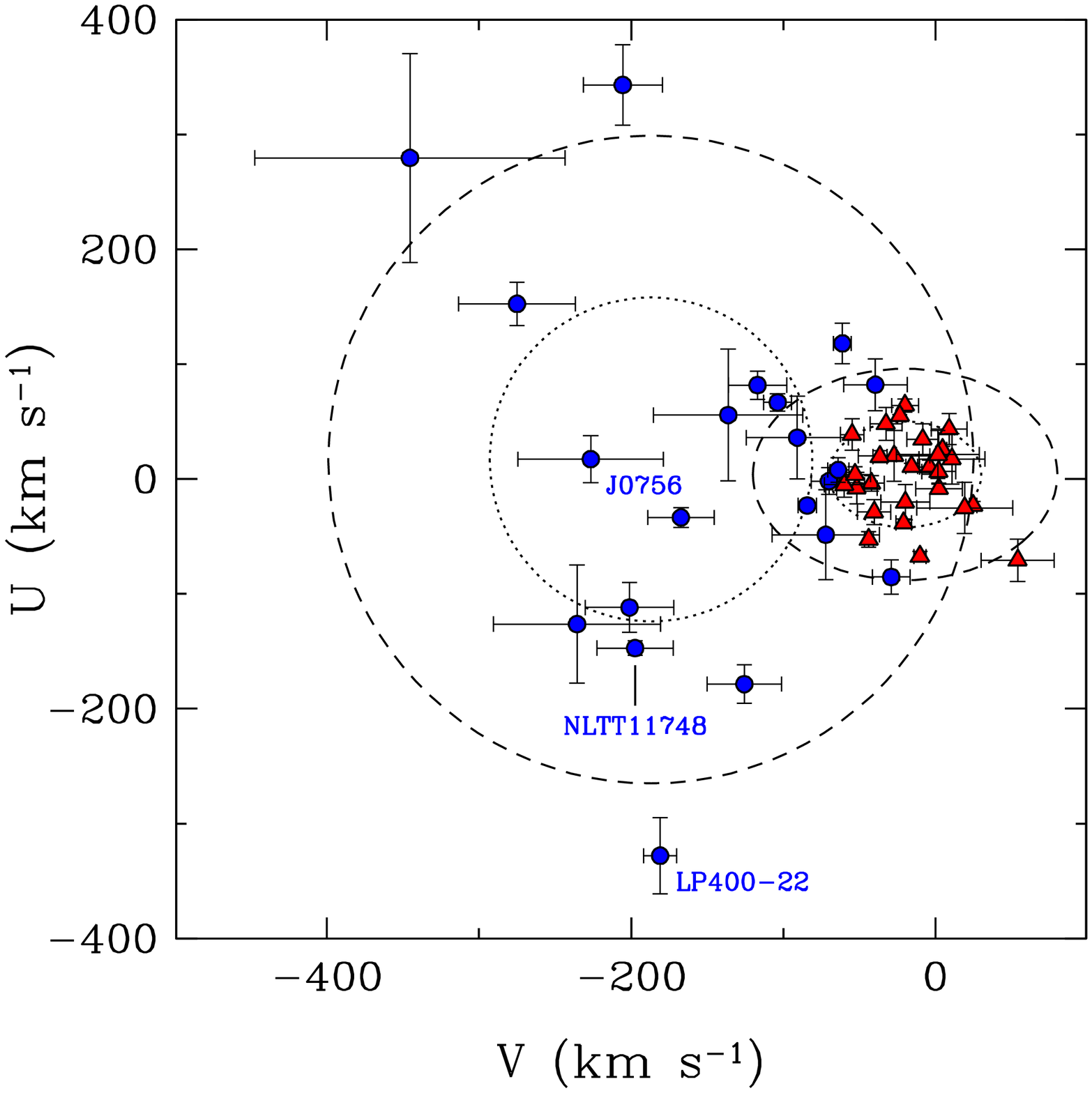}
\figcaption[f11a.eps]{$W$ vs. $U$ (top) and $U$ vs. $V$ (bottom)
  velocity distributions for 49 ELM WD binaries including nine
  from this paper. Blue circles denote ELM WDs with a Mahalanobis
  distance suggesting halo membership while the red triangles
  represent ELM WDs with kinematics consistent with the Galactic
  thick-disk.  The ellipsoids denote the 1-$\sigma$ (dotted) and
  2-$\sigma$ (dashed) contours for Galactic thick-disk and stellar
  halo populations.
\label{fg:UVW}}
\end{figure}

In Table~\ref{tab:UVW} we list the ($U$, $V$, $W$) space velocities
for 49 of the 73 WDs from the ELM Survey for which proper
motions were available from the SDSS+USNO-B catalog of \citet{munn04}.
In addition, the proper motion for J0345+1748 (\nltt) is taken from
\citet{kawka09} and its systemic velocity from \citet{kilic10}.
Finally we adopt the proper motion from UCAC4 for J0308+5140.

We computed the ($U$, $V$, $W$) space velocities and their associated
uncertainties according to the prescription of \citet{johnson87}
combining the proper motions with our determinations of the systemic
velocity (see Table~\ref{tab:bin}). It is important to note that since
we do not have parallax measurements for our targets, we adopt our
spectroscopically determined distances as a proxy for the parallax.
In addition, we correct for the motion of the local standard of rest
using ($U$, $V$, $W$)$_{\odot}$~=~(11.10, 12.24, 7.25)~\kms\ as
determined by \citet{schonrich10}.

In Figure~\ref{fg:UVW} we plot the $W$ vs. $V$ and $U$ vs. $V$
velocity distributions. In addition, we plot ellipsoids representing
Galactic thick-disk and stellar halo populations. We adopt average
velocities of ($\langle U \rangle$, $\langle V \rangle$, $\langle W
\rangle$)~=~(4, $-$20, $-$3)~\kms\ and (17, $-$187, $-5$)~\kms, and
velocity dispersions of ($\sigma_{U}$, $\sigma_{V}$,
$\sigma_{W}$)~=~(46, 50, 35)~\kms\ and (141, 106, 94)~\kms\ for the
thick-disk and halo, respectively \citep{chiba00}.

Treating the ellipsoids for the thick-disk and halo as multivariate
normal distributions, we compute the Mahalanobis distance, as defined
in Equation (\ref{mahala}), for each ELM WD. This measures the
distance from the center of the distributions in units of standard
deviations and provides a quantitative indicator of group membership.

\begin{equation}
D_{\rm m} = \sqrt{\frac{(U-\langle U \rangle)^{2}}{\sigma_{U}^{2}} + \frac{(V-\langle V \rangle)^{2}}{\sigma_{V}^{2}} + \frac{(W-\langle W \rangle)^{2}}{\sigma_{W}^{2}}} 
\label{mahala}
\end{equation}

\noindent The Mahalanobis distances with respect to the thick-disk and
halo distributions, $D_{\rm m, disk}$ and $D_{\rm m, halo}$, are
listed in Table \ref{tab:UVW}. Based on these values, 27 ELM WDs
have distances consistent with disk membership while the remaining 22
WDs have distances consistent with halo membership. J0756+6704, the
ELM WD with the most negative systemic velocity among the 11 new
ELM WD binaries, numbers among the likely halo
members. Unsurprisingly, \lp\ (J2236+2232) also stands out in the
bottom panel of Figure~\ref{fg:UVW} with $U$~=~$-$332~\kms. The case
of \lp\ is well documented \citep{kawka06,vennes09,kilic13b} and the
results here serve only to further confirm that \lp\ is indeed a
unique member of the halo.  We also find that \nltt\ is a halo object,
in agreement with \citet{kawka09} and \citet{kilic10}.

The breakdown of halo versus thick-disk members suggests that
$\approx$~40\% of the WDs from the ELM Survey are members of the
Galactic halo. However, this result is based solely on a kinematic
metric. Additional considerations, such as the location of the ELM WDs
versus the scale height of the disk, may provide additional
information not considered here. As such, we suggest that our result
represents an upper limit for the percentage of halo objects in the
ELM Survey.

Having said that, the fact that our sample includes a significant
fraction of halo stars should not be unexpected. This result is most
likely a logical consequence of our candidates being largely selected
from the SDSS which is a high Galactic latitude survey. Given the
luminosity of our targets, our apparent magnitude range samples white
dwarfs in the range $0.1 \lesssim |Z| \lesssim 2$~kpc, where $Z$
denotes the vertical distance from the Galactic plane. Given the scale
height of the disk and the relative normalization of the disk and
halo, it is not surprising that the ELM Survey contains a significant
number of halo stars.

\section{CONCLUSIONS}

We present radial velocity measurements and stellar atmosphere fits
for 11 new ELM WD binaries with $P < 1$~d. This brings the total
of ELM WDs identified by the ELM Survey up to 73 demonstrating
the continued effectiveness of using SDSS colors to identify ELM WD
candidates. Based our optical spectra, we perform spectroscopic fits,
compute orbital solutions, and provide a complete set of physical and
binary parameters for each system.

We review the recent evolutionary calculations of \citet{istrate14}
and compare them to the existing models of \citet{althaus13}. There
are significant differences between the two sets of evolutionary
models. These can be traced back to the different initial assumptions
made about the nature and composition of the progenitor binaries. We
can only conclude that there still remains much uncertainty regarding
the exact evolutionary history of ELM WDs and additional modeling is
required to form a coherent picture. Thankfully, the systematic
uncertainty when deriving masses from the independent model sets is
only $\approx$~0.01~\msun.

When considering the distribution of orbital periods as a function of
\Te, we have shown that two distinct populations are emerging. The
shorter period systems are generally found with \Te~$>$~15,000~K while
the longer period systems can cool to much lower \Te.

Of the 11 new ELM WD binaries, five will merge within a Hubble
time. Unfortunately, the final merger product for the majority of ELM
WD binaries remains uncertain. Nonetheless, we have identified four
systems which will undergo stable mass transfer and ultimately become
AM CVn systems.

Based on our updated boundaries of the instability strip, we have also
identified a number of potentially interesting targets for follow-up
high-speed photometric observations. These ELM WDs have atmospheric
parameters within or near the instability strip of pulsating ELM WDs,
an extension of the canonical ZZ Ceti instability strip.

Finally, a study of the kinematics of ELM WDs reveals that the
majority of these binaries are members of the Galactic disk. However,
a non-negligible fraction of ELM WDs are most certainly members of the
Galactic halo, a result of our reliance on SDSS colors to identify
candidates.

\acknowledgements We would like to thank the referee and statistics
consultant for their useful suggestions which helped improve this
paper. We also thank the director and staff of Kitt Peak National
Observatory for the use of their facilities and their valuable
assistance. We gratefully acknowledge the support of the NSF under
grant AST-1312678, and NASA under grant NNX14AF65G.  This work was
supported in part by National Science Foundation Grant
No. PHYS-1066293 and the hospitality of the Aspen Center for Physics.
AG acknowledges support provided by NASA through grant number
HST-GO-13319.01 from the Space Telescope Science Institute, which is
operated by AURA, Inc., under NASA contract NAS 5-26555. This material
is based upon work supported by AURA through the National Science
Foundation under AURA Cooperative Agreement AST 0132798 as amended.

{\it Facilities:} \facility{MMT (Blue Channel Spectrograph), KPNO 4m
  telescope (RCSpec, KOSMOS), Palomar Hale 5m telescope (Double Spec), FWLO
  1.5m telescope (FAST)}

\appendix

\begin{table*}
\scriptsize
\caption{ELM WD Physical Parameters}
\begin{center}
\setlength{\tabcolsep}{7.2pt}
\begin{tabular*}{\hsize}{@{\extracolsep{\fill}}lr@{ $\pm$ }@{\extracolsep{0pt}}lr@{ $\pm$ }@{\extracolsep{0pt}}lr@{ $\pm$ }@{\extracolsep{0pt}}lr@{ $\pm$ }@{\extracolsep{0pt}}lr@{ $\pm$ }@{\extracolsep{0pt}}lr@{ $\pm$ }@{\extracolsep{0pt}}lr@{ $\pm$ }@{\extracolsep{0pt}}lr@{ $\pm$ }@{\extracolsep{0pt}}l@{}}
\hline
\hline
\noalign{\smallskip}
SDSS & \multicolumn{2}{c}{\Te} & \multicolumn{2}{c}{\logg}        & \multicolumn{2}{c}{$M_{1}$}  & \multicolumn{2}{c}{Radius}  & \multicolumn{2}{c}{$g_{0}$} & \multicolumn{2}{c}{$M_{g}$} & \multicolumn{2}{c}{$d$}  & \multicolumn{2}{c}{$\tau_{\rm cool}$} \\
     & \multicolumn{2}{c}{(K)} & \multicolumn{2}{c}{(cm s$^{-2}$)} & \multicolumn{2}{c}{(\msun)} & \multicolumn{2}{c}{(\rsun)} & \multicolumn{2}{c}{(mag)}  & \multicolumn{2}{c}{(mag)}  & \multicolumn{2}{c}{(kpc)} & \multicolumn{2}{c}{(Gyr)}          \\
\noalign{\smallskip}
\hline
\noalign{\smallskip}
J0022+0031              & 20460 & 310 & 7.58 & 0.05 & 0.457 & 0.016 & 0.0182 & 0.0013 & 19.284 & 0.033 &  9.84 & 0.18 & 0.774 & 0.065 & 0.215 & 0.128 \\
J0022$-$1014            & 20730 & 340 & 7.28 & 0.05 & 0.375 & 0.016 & 0.0233 & 0.0018 & 19.581 & 0.031 &  9.28 & 0.20 & 1.151 & 0.107 & 0.042 & 0.021 \\
J0056$-$0611            & 12240 & 180 & 6.18 & 0.04 & 0.174 & 0.010 & 0.0564 & 0.0045 & 17.208 & 0.023 &  8.37 & 0.21 & 0.585 & 0.056 & 0.957 & 0.081 \\
J0106$-$1000            & 16970 & 260 & 6.10 & 0.05 & 0.191 & 0.010 & 0.0642 & 0.0051 & 19.595 & 0.023 &  7.45 & 0.20 & 2.691 & 0.253 & 0.497 & 0.168 \\
J0112+1835              &  9740 & 140 & 5.77 & 0.05 & 0.160 & 0.010 & 0.0863 & 0.0080 & 17.110 & 0.016 &  8.01 & 0.25 & 0.662 & 0.077 & 1.822 & 0.196 \\
J0152+0749              & 10800 & 180 & 5.93 & 0.05 & 0.168 & 0.010 & 0.0736 & 0.0061 & 18.033 & 0.009 &  8.08 & 0.22 & 0.980 & 0.102 & 1.407 & 0.137 \\
\noalign{\smallskip}
\hline
\noalign{\smallskip}
\multicolumn{17}{@{}l}{(This table is available in its entirety in a machine-readable form in the online journal. A portion is shown here for guidance regarding its form and content.)}
\label{tab:par2}
\end{tabular*}
\end{center}

\caption{ELM WD Binary Parameters}
\begin{center}
\setlength{\tabcolsep}{8.15pt}
\begin{tabular*}{\hsize}{@{\extracolsep{\fill}}lr@{ $\pm$ }@{\extracolsep{0pt}}lr@{ $\pm$ }@{\extracolsep{0pt}}lr@{ $\pm$ }@{\extracolsep{0pt}}lr@{ $\pm$ }@{\extracolsep{0pt}}lr@{ $\pm$ }@{\extracolsep{0pt}}lcr@{ $\pm$ }@{\extracolsep{0pt}}lc@{}}
\hline
\hline
\noalign{\smallskip}
SDSS & \multicolumn{2}{c}{$P$} & \multicolumn{2}{c}{$K$} & \multicolumn{2}{c}{Mass Function} & \multicolumn{2}{c}{$M_{2}$} & \multicolumn{2}{c}{$M_{2, i=60^{\circ}}$} & $\tau_{\rm merge}$ & \multicolumn{2}{c}{$a$} & $\log h$ \\
     & \multicolumn{2}{c}{(days)}       & \multicolumn{2}{c}{(\kms)} & \multicolumn{2}{c}{(\msun)} & \multicolumn{2}{c}{(\msun)} & \multicolumn{2}{c}{(\msun)} & (Gyr) & \multicolumn{2}{c}{(\rsun)} & \\
\noalign{\smallskip}
\hline
\noalign{\smallskip}
J0022+0031   & 0.49135 & 0.02540 &  80.8 & 1.3 & 0.027 & 0.003 & $\geqslant$ 0.23 & 0.02 & 0.28 & 0.02 & \ldots            & 2.32 & 0.11 & $-$22.79 \\
J0022$-$1014 & 0.07989 & 0.00300 & 145.6 & 5.6 & 0.026 & 0.004 & $\geqslant$ 0.21 & 0.02 & 0.25 & 0.02 & $\leqslant$ 0.616 & 0.65 & 0.03 & $-$22.55 \\
J0056$-$0611 & 0.04338 & 0.00002 & 376.9 & 2.4 & 0.241 & 0.005 & $\geqslant$ 0.46 & 0.02 & 0.61 & 0.02 & $\leqslant$ 0.120 & 0.45 & 0.01 & $-$22.06 \\
J0106$-$1000 & 0.02715 & 0.00002 & 395.2 & 3.6 & 0.174 & 0.005 & $\geqslant$ 0.39 & 0.02 & 0.51 & 0.02 & $\leqslant$ 0.036 & 0.32 & 0.01 & $-$22.61 \\
J0112+1835   & 0.14698 & 0.00003 & 295.3 & 2.0 & 0.392 & 0.008 & $\geqslant$ 0.62 & 0.02 & 0.85 & 0.03 & $\leqslant$ 2.682 & 1.08 & 0.01 & $-$22.40 \\
J0152+0749   & 0.32288 & 0.00014 & 217.0 & 2.0 & 0.342 & 0.010 & $\geqslant$ 0.57 & 0.02 & 0.78 & 0.03 & \ldots            & 1.79 & 0.03 & $-$22.80 \\
\noalign{\smallskip}
\hline
\noalign{\smallskip}
\multicolumn{15}{@{}l}{(This table is available in its entirety in a machine-readable form in the online journal. A portion is shown here for guidance regarding its form and content.)}
\end{tabular*}
\label{tab:bin2}
\end{center}
\end{table*}

\begin{table}[!h]
\caption{Radial Velocity Measurements}
\begin{center}

\begin{tabular}{@{\extracolsep{\fill}}l@{\extracolsep{46pt}}cr@{ $\pm$ }@{\extracolsep{0pt}}r@{}}
\hline
\hline
\noalign{\smallskip}
Object & HJD                & \multicolumn{2}{c}{$v_{\rm helio}$} \\
       & (days~$-$~2455000) & \multicolumn{2}{c}{(\kms)}       \\
\noalign{\smallskip}
\hline
\noalign{\smallskip}
J0308+5140   & 1634.76726 &     19.2 & 13.6 \\
\ldots       & 1716.63827 & $-$145.6 & 19.9 \\
\ldots       & 1716.64861 & $-$132.3 & 22.6 \\
\ldots       & 1716.65891 & $-$154.4 & 22.8 \\
\ldots       & 1716.66922 & $-$134.4 & 26.4 \\
\ldots       & 1946.81093 &      9.7 & 12.1 \\ 
\noalign{\smallskip}
\hline
\noalign{\smallskip}
\multicolumn{4}{@{}p{\hsize}@{}}{(This table is available in its entirety in a machine-readable form in the online journal. A portion is shown here for guidance regarding its form and content.)}
\end{tabular}
\end{center}
\label{tab:data}
\end{table}

\section{ELM WD Parameters}

We provide in Tables~\ref{tab:par2} and \ref{tab:bin2} a complete
listing of the physical and binary parameters for the entire ELM
Survey sample. This includes the 62 ELM WDs from previous publications
as well as the 11 new ELM WDs presented here. We note that for
ELM WDs with \Te~$<$ 12,000~K, the atmospheric parameters have been
adjusted using the 3D corrections from \citet{tremblay15}. We also
note that we have updated our value of the apparent magnitude for
J0345+1748 (\nltt).  In \citet{gianninas14b} we had adopted the value
of $V$~=~16.5 from \citet{kawka09}. We now adopt the value of
$g$~=~16.797 from UCAC4.  We then take $A_{g}/E(B-V)$~=~3.793 from
IRSA and $E(B-V)$~=~0.10 from \citet{kawka09} to obtain an extinction
of $A_{g}$~=~0.3793. Thus, our final extinction-corrected $g$-band
magnitude for \nltt\ is $g_{0}$~=~16.418.

\section{DATA TABLE}

Table~\ref{tab:data} presents our radial velocity measurements for the
11 new ELM WD binaries presented here. The table columns include
object name, heliocentric Julian date, heliocentric radial velocity,
and velocity error.

\break

\bibliographystyle{apj}
\bibliography{biblio}

\begin{thebibliography}{}
\expandafter\ifx\csname natexlab\endcsname\relax\def\natexlab#1{#1}\fi

\bibitem[{{Althaus} {et~al.}(2013){Althaus}, {Miller Bertolami}, \&
  {C{\'o}rsico}}]{althaus13}
{Althaus}, L.~G., {Miller Bertolami}, M.~M., \& {C{\'o}rsico}, A.~H. 2013,
  \aap, 557, A19

\bibitem[{{Amaro-Seoane} {et~al.}(2012){Amaro-Seoane}, {Aoudia}, {Babak},
  {Bin{\'e}truy}, {Berti}, {Boh{\'e}}, {Caprini}, {Colpi}, {Cornish},
  {Danzmann}, {Dufaux}, {Gair}, {Jennrich}, {Jetzer}, {Klein}, {Lang}, {Lobo},
  {Littenberg}, {McWilliams}, {Nelemans}, {Petiteau}, {Porter}, {Schutz},
  {Sesana}, {Stebbins}, {Sumner}, {Vallisneri}, {Vitale}, {Volonteri}, \&
  {Ward}}]{amaro12}
{Amaro-Seoane}, P., {Aoudia}, S., {Babak}, S., {et~al.} 2012, Classical and
  Quantum Gravity, 29, 124016

\bibitem[{{Andrews} {et~al.}(2014){Andrews}, {Price-Whelan}, \&
  {Ag{\"u}eros}}]{andrews14}
{Andrews}, J.~J., {Price-Whelan}, A.~M., \& {Ag{\"u}eros}, M.~A. 2014, \apjl,
  797, L32

\bibitem[{{Antoniadis} {et~al.}(2012){Antoniadis}, {van Kerkwijk}, {Koester},
  {Freire}, {Wex}, {Tauris}, {Kramer}, \& {Bassa}}]{antoniadis12}
{Antoniadis}, J., {van Kerkwijk}, M.~H., {Koester}, D., {et~al.} 2012, \mnras,
  423, 3316

\bibitem[{{Bildsten} {et~al.}(2007){Bildsten}, {Shen}, {Weinberg}, \&
  {Nelemans}}]{bildsten07}
{Bildsten}, L., {Shen}, K.~J., {Weinberg}, N.~N., \& {Nelemans}, G. 2007,
  \apjl, 662, L95

\bibitem[{{Boffin}(2015)}]{boffin15}
{Boffin}, H.~M.~J. 2015, \aap, 575, L13

\bibitem[{{Brown} {et~al.}(2011{\natexlab{a}}){Brown}, {Kilic}, {Brown}, \&
  {Kenyon}}]{brownj11}
{Brown}, J.~M., {Kilic}, M., {Brown}, W.~R., \& {Kenyon}, S.~J.
  2011{\natexlab{a}}, \apj, 730, 67

\bibitem[{{Brown} {et~al.}(2013){Brown}, {Kilic}, {Allende Prieto},
  {Gianninas}, \& {Kenyon}}]{brown_ELM5}
{Brown}, W.~R., {Kilic}, M., {Allende Prieto}, C., {Gianninas}, A., \&
  {Kenyon}, S.~J. 2013, \apj, 769, 66

\bibitem[{{Brown} {et~al.}(2010){Brown}, {Kilic}, {Allende Prieto}, \&
  {Kenyon}}]{brown_ELM1}
{Brown}, W.~R., {Kilic}, M., {Allende Prieto}, C., \& {Kenyon}, S.~J. 2010,
  \apj, 723, 1072

\bibitem[{{Brown} {et~al.}(2012){Brown}, {Kilic}, {Allende Prieto}, \&
  {Kenyon}}]{brown_ELM3}
---. 2012, \apj, 744, 142

\bibitem[{{Brown} {et~al.}(2011{\natexlab{b}}){Brown}, {Kilic}, {Hermes},
  {Allende Prieto}, {Kenyon}, \& {Winget}}]{brown11}
{Brown}, W.~R., {Kilic}, M., {Hermes}, J.~J., {et~al.} 2011{\natexlab{b}},
  \apjl, 737, L23

\bibitem[{{Chiba} \& {Beers}(2000)}]{chiba00}
{Chiba}, M., \& {Beers}, T.~C. 2000, \aj, 119, 2843

\bibitem[{{Clayton}(2013)}]{clayton13}
{Clayton}, G.~C. 2013, in Astronomical Society of the Pacific Conference
  Series, Vol. 469, 18th European White Dwarf Workshop., ed. {Krzesi{\'n}},
  J.~{ski}, G.~{Stachowski}, P.~{Moskalik}, \& K.~{Bajan}, 133

\bibitem[{{Cui} {et~al.}(2012){Cui}, {Zhao}, {Chu}, {Li}, {Li}, {Zhang}, {Su},
  {Yao}, {Wang}, {Xing}, {Li}, {Zhu}, {Wang}, {Gu}, {Luo}, {Xu}, {Zhang},
  {Liu}, {Zhang}, {Yang}, {Cao}, {Chen}, {Chen}, {Chen}, {Chen}, {Chu}, {Feng},
  {Gong}, {Hou}, {Hu}, {Hu}, {Hu}, {Jia}, {Jiang}, {Jiang}, {Jiang}, {Jin},
  {Li}, {Li}, {Li}, {Liu}, {Liu}, {Lu}, {Mao}, {Men}, {Qi}, {Qi}, {Shi},
  {Tang}, {Tao}, {Wang}, {Wang}, {Wang}, {Wang}, {Wang}, {Wang}, {Wang},
  {Wang}, {Wang}, {Wang}, {Wang}, {Wang}, {Xu}, {Xu}, {Yang}, {Yu}, {Yuan},
  {Yuan}, {Zhai}, {Zhang}, {Zhang}, {Zhang}, {Zhao}, {Zhou}, {Zhou}, {Zhu}, \&
  {Zou}}]{cui12}
{Cui}, X.-Q., {Zhao}, Y.-H., {Chu}, Y.-Q., {et~al.} 2012, Research in Astronomy
  and Astrophysics, 12, 1197

\bibitem[{{Debes} {et~al.}(2015){Debes}, {Kilic}, {Tremblay},
  {L{\'o}pez-Morales}, {Anglada-Escude}, {Napiwotzki}, {Osip}, \&
  {Weinberger}}]{debes15}
{Debes}, J.~H., {Kilic}, M., {Tremblay}, P.-E., {et~al.} 2015, \aj, 149, 176

\bibitem[{{Fabricant} {et~al.}(1998){Fabricant}, {Cheimets}, {Caldwell}, \&
  {Geary}}]{fabricant98}
{Fabricant}, D., {Cheimets}, P., {Caldwell}, N., \& {Geary}, J. 1998, \pasp,
  110, 79

\bibitem[{{Foley}(2015)}]{foley15}
{Foley}, R.~J. 2015, ArXiv e-prints, arXiv:1501.07607

\bibitem[{{Gianninas} {et~al.}(2011){Gianninas}, {Bergeron}, \&
  {Ruiz}}]{gianninas11}
{Gianninas}, A., {Bergeron}, P., \& {Ruiz}, M.~T. 2011, \apj, 743, 138

\bibitem[{{Gianninas} {et~al.}(2014{\natexlab{a}}){Gianninas}, {Dufour},
  {Kilic}, {Brown}, {Bergeron}, \& {Hermes}}]{gianninas14a}
{Gianninas}, A., {Dufour}, P., {Kilic}, M., {et~al.} 2014{\natexlab{a}}, \apj,
  794, 35

\bibitem[{{Gianninas} {et~al.}(2014{\natexlab{b}}){Gianninas}, {Hermes},
  {Brown}, {Dufour}, {Barber}, {Kilic}, {Kenyon}, \& {Harrold}}]{gianninas14b}
{Gianninas}, A., {Hermes}, J.~J., {Brown}, W.~R., {et~al.} 2014{\natexlab{b}},
  \apj, 781, 104

\bibitem[{{Green} {et~al.}(2015){Green}, {Limoges}, {Gianninas}, {Bergeron},
  {Fontaine}, {Dufour}, {O'Malley}, {Guvenen}, {Biddle}, {Pearson}, {Deyoe},
  {Bullivant}, {Hermes}, {Van Grootel}, \& {Grosjean}}]{green15}
{Green}, E.~M., {Limoges}, M.-M., {Gianninas}, A., {et~al.} 2015, in
  Astronomical Society of the Pacific Conference Series, Vol. 493, 19th
  European Workshop on White Dwarfs, ed. P.~{Dufour}, P.~{Bergeron}, \&
  G.~{Fontaine}, 237

\bibitem[{{Hallakoun} {et~al.}(2015){Hallakoun}, {Maoz}, {Kilic}, {Mazeh},
  {Agol}, {Bell}, {Bloemen}, {Brown}, {Debes}, {Faigler}, {Gianninas}, {Kull},
  {Kupfer}, {Loeb}, {Morris}, \& {Mullally}}]{hallakoun15}
{Hallakoun}, N., {Maoz}, D., {Kilic}, M., {et~al.} 2015, ArXiv e-prints,
  arXiv:1507.06311

\bibitem[{{Hansen} {et~al.}(2007){Hansen}, {Anderson}, {Brewer}, {Dotter},
  {Fahlman}, {Hurley}, {Kalirai}, {King}, {Reitzel}, {Richer}, {Rich}, {Shara},
  \& {Stetson}}]{hansen07}
{Hansen}, B.~M.~S., {Anderson}, J., {Brewer}, J., {et~al.} 2007, \apj, 671, 380

\bibitem[{{Hermes} {et~al.}(2013{\natexlab{a}}){Hermes}, {Montgomery},
  {Gianninas}, {Winget}, {Brown}, {Harrold}, {Bell}, {Kenyon}, {Kilic}, \&
  {Castanheira}}]{hermes13c}
{Hermes}, J.~J., {Montgomery}, M.~H., {Gianninas}, A., {et~al.}
  2013{\natexlab{a}}, \mnras, 436, 3573

\bibitem[{{Hermes} {et~al.}(2013{\natexlab{b}}){Hermes}, {Montgomery},
  {Winget}, {Brown}, {Gianninas}, {Kilic}, {Kenyon}, {Bell}, \&
  {Harrold}}]{hermes13b}
{Hermes}, J.~J., {Montgomery}, M.~H., {Winget}, D.~E., {et~al.}
  2013{\natexlab{b}}, \apj, 765, 102

\bibitem[{{Hermes} {et~al.}(2014){Hermes}, {Brown}, {Kilic}, {Gianninas},
  {Chote}, {Sullivan}, {Winget}, {Bell}, {Falcon}, {Winget}, {Mason},
  {Harrold}, \& {Montgomery}}]{hermes14}
{Hermes}, J.~J., {Brown}, W.~R., {Kilic}, M., {et~al.} 2014, \apj, 792, 39

\bibitem[{{Iben} \& {Tutukov}(1984)}]{iben84}
{Iben}, Jr., I., \& {Tutukov}, A.~V. 1984, \apjs, 54, 335

\bibitem[{{Istrate} {et~al.}(2014){Istrate}, {Tauris}, {Langer}, \&
  {Antoniadis}}]{istrate14}
{Istrate}, A.~G., {Tauris}, T.~M., {Langer}, N., \& {Antoniadis}, J. 2014,
  \aap, 571, L3

\bibitem[{{Johnson} \& {Soderblom}(1987)}]{johnson87}
{Johnson}, D.~R.~H., \& {Soderblom}, D.~R. 1987, \aj, 93, 864

\bibitem[{{Kawka} \& {Vennes}(2009)}]{kawka09}
{Kawka}, A., \& {Vennes}, S. 2009, \aap, 506, L25

\bibitem[{{Kawka} {et~al.}(2006){Kawka}, {Vennes}, {Oswalt}, {Smith}, \&
  {Silvestri}}]{kawka06}
{Kawka}, A., {Vennes}, S., {Oswalt}, T.~D., {Smith}, J.~A., \& {Silvestri},
  N.~M. 2006, \apjl, 643, L123

\bibitem[{{Kenyon} \& {Garcia}(1986)}]{kenyon86}
{Kenyon}, S.~J., \& {Garcia}, M.~R. 1986, \aj, 91, 125

\bibitem[{{Kilic} {et~al.}(2010{\natexlab{a}}){Kilic}, {Allende Prieto},
  {Brown}, {Ag{\"u}eros}, {Kenyon}, \& {Camilo}}]{kilic10}
{Kilic}, M., {Allende Prieto}, C., {Brown}, W.~R., {et~al.} 2010{\natexlab{a}},
  \apjl, 721, L158

\bibitem[{{Kilic} {et~al.}(2011){Kilic}, {Brown}, {Allende Prieto},
  {Ag{\"u}eros}, {Heinke}, \& {Kenyon}}]{kilic_ELM2}
{Kilic}, M., {Brown}, W.~R., {Allende Prieto}, C., {et~al.} 2011, \apj, 727, 3

\bibitem[{{Kilic} {et~al.}(2012){Kilic}, {Brown}, {Allende Prieto}, {Kenyon},
  {Heinke}, {Ag{\"u}eros}, \& {Kleinman}}]{kilic_ELM4}
---. 2012, \apj, 751, 141

\bibitem[{{Kilic} {et~al.}(2010{\natexlab{b}}){Kilic}, {Brown}, {Allende
  Prieto}, {Kenyon}, \& {Panei}}]{kilic_ELM0}
{Kilic}, M., {Brown}, W.~R., {Allende Prieto}, C., {Kenyon}, S.~J., \& {Panei},
  J.~A. 2010{\natexlab{b}}, \apj, 716, 122

\bibitem[{{Kilic} {et~al.}(2014{\natexlab{a}}){Kilic}, {Brown}, {Gianninas},
  {Hermes}, {Allende Prieto}, \& {Kenyon}}]{kilic14a}
{Kilic}, M., {Brown}, W.~R., {Gianninas}, A., {et~al.} 2014{\natexlab{a}},
  \mnras, 444, L1

\bibitem[{{Kilic} {et~al.}(2015){Kilic}, {Hermes}, {Gianninas}, \&
  {Brown}}]{kilic15}
{Kilic}, M., {Hermes}, J.~J., {Gianninas}, A., \& {Brown}, W.~R. 2015, \mnras,
  446, L26

\bibitem[{{Kilic} {et~al.}(2013){Kilic}, {Gianninas}, {Brown}, {Harris},
  {Dahn}, {Ag{\"u}eros}, {Heinke}, {Kenyon}, {Panei}, \& {Camilo}}]{kilic13b}
{Kilic}, M., {Gianninas}, A., {Brown}, W.~R., {et~al.} 2013, \mnras, 434, 3582

\bibitem[{{Kilic} {et~al.}(2014{\natexlab{b}}){Kilic}, {Hermes}, {Gianninas},
  {Brown}, {Heinke}, {Ag{\"u}eros}, {Chote}, {Sullivan}, {Bell}, \&
  {Harrold}}]{kilic14b}
{Kilic}, M., {Hermes}, J.~J., {Gianninas}, A., {et~al.} 2014{\natexlab{b}},
  \mnras, 438, L26

\bibitem[{{Kremer} {et~al.}(2015){Kremer}, {Sepinsky}, \&
  {Kalogera}}]{kremer15}
{Kremer}, K., {Sepinsky}, J., \& {Kalogera}, V. 2015, \apj, 806, 76

\bibitem[{{Kurtz} \& {Mink}(1998)}]{kurtz98}
{Kurtz}, M.~J., \& {Mink}, D.~J. 1998, \pasp, 110, 934

\bibitem[{{Landau} \& {Lifshitz}(1958)}]{landau58}
{Landau}, L.~D., \& {Lifshitz}, E.~M. 1958, {The Classical Theory of Fields}
  (Oxford: Pergamon Press)

\bibitem[{{Liebert} {et~al.}(2005){Liebert}, {Bergeron}, \& {Holberg}}]{LBH05}
{Liebert}, J., {Bergeron}, P., \& {Holberg}, J.~B. 2005, \apjs, 156, 47

\bibitem[{{Marsh} {et~al.}(1995){Marsh}, {Dhillon}, \& {Duck}}]{marsh95}
{Marsh}, T.~R., {Dhillon}, V.~S., \& {Duck}, S.~R. 1995, \mnras, 275, 828

\bibitem[{{Marsh} {et~al.}(2004){Marsh}, {Nelemans}, \& {Steeghs}}]{marsh04}
{Marsh}, T.~R., {Nelemans}, G., \& {Steeghs}, D. 2004, \mnras, 350, 113

\bibitem[{{Martini} {et~al.}(2014){Martini}, {Elias}, {Points}, {Sprayberry},
  {Derwent}, {Gonzalez}, {Mason}, {O'Brien}, {Pappalardo}, {Pogge}, {Stoll},
  {Zhelem}, {Daly}, {Fitzpatrick}, {George}, {Hunten}, {Marshall}, {Poczulp},
  {Rath}, {Seaman}, {Trueblood}, \& {Zelaya}}]{martini14}
{Martini}, P., {Elias}, J., {Points}, S., {et~al.} 2014, in Society of
  Photo-Optical Instrumentation Engineers (SPIE) Conference Series, Vol. 9147,
  Society of Photo-Optical Instrumentation Engineers (SPIE) Conference Series,
  91470Z

\bibitem[{{Massey} {et~al.}(1988){Massey}, {Strobel}, {Barnes}, \&
  {Anderson}}]{massey88}
{Massey}, P., {Strobel}, K., {Barnes}, J.~V., \& {Anderson}, E. 1988, \apj,
  328, 315

\bibitem[{{Moran} {et~al.}(1997){Moran}, {Marsh}, \& {Bragaglia}}]{moran97}
{Moran}, C., {Marsh}, T.~R., \& {Bragaglia}, A. 1997, \mnras, 288, 538

\bibitem[{{Munn} {et~al.}(2004){Munn}, {Monet}, {Levine}, {Canzian}, {Pier},
  {Harris}, {Lupton}, {Ivezi{\'c}}, {Hindsley}, {Hennessy}, {Schneider}, \&
  {Brinkmann}}]{munn04}
{Munn}, J.~A., {Monet}, D.~G., {Levine}, S.~E., {et~al.} 2004, \aj, 127, 3034

\bibitem[{{Napiwotzki} {et~al.}(2007){Napiwotzki}, {Karl}, {Nelemans},
  {Yungelson}, {Christlieb}, {Drechsel}, {Heber}, {Homeier}, {Koester},
  {Leibundgut}, {Marsh}, {Moehler}, {Renzini}, \& {Reimers}}]{napiwotzki07}
{Napiwotzki}, R., {Karl}, C.~A., {Nelemans}, G., {et~al.} 2007, in Astronomical
  Society of the Pacific Conference Series, Vol. 372, 15th European Workshop on
  White Dwarfs, ed. R.~{Napiwotzki} \& M.~R. {Burleigh}, 387

\bibitem[{{Nelson} {et~al.}(2004){Nelson}, {Dubeau}, \&
  {MacCannell}}]{nelson04}
{Nelson}, L.~A., {Dubeau}, E., \& {MacCannell}, K.~A. 2004, \apj, 616, 1124

\bibitem[{{Oke} \& {Gunn}(1982)}]{oke82}
{Oke}, J.~B., \& {Gunn}, J.~E. 1982, \pasp, 94, 586

\bibitem[{{Parsons} {et~al.}(2011){Parsons}, {Marsh}, {G{\"a}nsicke}, {Drake},
  \& {Koester}}]{parsons11}
{Parsons}, S.~G., {Marsh}, T.~R., {G{\"a}nsicke}, B.~T., {Drake}, A.~J., \&
  {Koester}, D. 2011, \apjl, 735, L30

\bibitem[{{Perets} {et~al.}(2010){Perets}, {Gal-Yam}, {Mazzali}, {Arnett},
  {Kagan}, {Filippenko}, {Li}, {Arcavi}, {Cenko}, {Fox}, {Leonard}, {Moon},
  {Sand}, {Soderberg}, {Anderson}, {James}, {Foley}, {Ganeshalingam}, {Ofek},
  {Bildsten}, {Nelemans}, {Shen}, {Weinberg}, {Metzger}, {Piro}, {Quataert},
  {Kiewe}, \& {Poznanski}}]{perets10}
{Perets}, H.~B., {Gal-Yam}, A., {Mazzali}, P.~A., {et~al.} 2010, \nat, 465, 322

\bibitem[{{Press} {et~al.}(1986){Press}, {Flannery}, \& {Teukolsky}}]{press86}
{Press}, W.~H., {Flannery}, B.~P., \& {Teukolsky}, S.~A. 1986, {Numerical
  recipes. The art of scientific computing}

\bibitem[{{Schmidt} {et~al.}(1989){Schmidt}, {Weymann}, \& {Foltz}}]{schmidt89}
{Schmidt}, G.~D., {Weymann}, R.~J., \& {Foltz}, C.~B. 1989, \pasp, 101, 713

\bibitem[{{Sch{\"o}nrich} {et~al.}(2010){Sch{\"o}nrich}, {Binney}, \&
  {Dehnen}}]{schonrich10}
{Sch{\"o}nrich}, R., {Binney}, J., \& {Dehnen}, W. 2010, \mnras, 403, 1829

\bibitem[{{Seaton}(1979)}]{seaton79}
{Seaton}, M.~J. 1979, \mnras, 187, 73P

\bibitem[{{Shen}(2015)}]{shen15}
{Shen}, K.~J. 2015, \apjl, 805, L6

\bibitem[{{Shporer} {et~al.}(2010){Shporer}, {Kaplan}, {Steinfadt}, {Bildsten},
  {Howell}, \& {Mazeh}}]{shporer10}
{Shporer}, A., {Kaplan}, D.~L., {Steinfadt}, J.~D.~R., {et~al.} 2010, \apjl,
  725, L200

\bibitem[{{Solheim}(2010)}]{solheim10}
{Solheim}, J.-E. 2010, \pasp, 122, 1133

\bibitem[{{Steinfadt} {et~al.}(2010){Steinfadt}, {Kaplan}, {Shporer},
  {Bildsten}, \& {Howell}}]{steinfadt10}
{Steinfadt}, J.~D.~R., {Kaplan}, D.~L., {Shporer}, A., {Bildsten}, L., \&
  {Howell}, S.~B. 2010, \apjl, 716, L146

\bibitem[{{Tremblay} \& {Bergeron}(2009)}]{TB09}
{Tremblay}, P.-E., \& {Bergeron}, P. 2009, \apj, 696, 1755

\bibitem[{{Tremblay} {et~al.}(2015){Tremblay}, {Gianninas}, {Kilic}, {Ludwig},
  {Steffen}, {Freytag}, \& {Hermes}}]{tremblay15}
{Tremblay}, P.-E., {Gianninas}, A., {Kilic}, M., {et~al.} 2015, \apj, 809, 148

\bibitem[{{Tremblay} {et~al.}(2013){Tremblay}, {Ludwig}, {Steffen}, \&
  {Freytag}}]{tremblay13b}
{Tremblay}, P.-E., {Ludwig}, H.-G., {Steffen}, M., \& {Freytag}, B. 2013, \aap,
  559, A104

\bibitem[{{Van Grootel} {et~al.}(2013){Van Grootel}, {Fontaine}, {Brassard}, \&
  {Dupret}}]{vg13}
{Van Grootel}, V., {Fontaine}, G., {Brassard}, P., \& {Dupret}, M.-A. 2013,
  \apj, 762, 57

\bibitem[{{Vennes} {et~al.}(2009){Vennes}, {Kawka}, {Vaccaro}, \&
  {Silvestri}}]{vennes09}
{Vennes}, S., {Kawka}, A., {Vaccaro}, T.~R., \& {Silvestri}, N.~M. 2009, \aap,
  507, 1613

\bibitem[{{Vennes} {et~al.}(2011){Vennes}, {Thorstensen}, {Kawka},
  {N{\'e}meth}, {Skinner}, {Pigulski}, {St{\c e}{\'s}licki}, {licki},
  {Ko{\l}aczkowski}, \& {{\'S}r{\'o}dka}}]{vennes11}
{Vennes}, S., {Thorstensen}, J.~R., {Kawka}, A., {et~al.} 2011, \apjl, 737, L16

\bibitem[{{Wang} {et~al.}(1996){Wang}, {Su}, {Chu}, {Cui}, \& {Wang}}]{wang96}
{Wang}, S.-G., {Su}, D.-Q., {Chu}, Y.-Q., {Cui}, X., \& {Wang}, Y.-N. 1996,
  \ao, 35, 5155

\bibitem[{{Webbink}(1984)}]{webbink84}
{Webbink}, R.~F. 1984, \apj, 277, 355

\bibitem[{{Zacharias} {et~al.}(2013){Zacharias}, {Finch}, {Girard}, {Henden},
  {Bartlett}, {Monet}, \& {Zacharias}}]{zacharias13}
{Zacharias}, N., {Finch}, C.~T., {Girard}, T.~M., {et~al.} 2013, \aj, 145, 44

\bibitem[{{Zhao} {et~al.}(2013){Zhao}, {Luo}, {Oswalt}, \& {Zhao}}]{zhao13}
{Zhao}, J.~K., {Luo}, A.~L., {Oswalt}, T.~D., \& {Zhao}, G. 2013, \aj, 145, 169

\end{thebibliography}

\end{document}